\documentclass[%
 aps,
 prb,%
 amsmath,amssymb,
reprint%
]{revtex4-2}

\usepackage{graphicx}
\usepackage{dcolumn}
\usepackage{bm}
\usepackage{bbm}
\usepackage{hyperref}
\usepackage{comment}
\usepackage{simpler-wick}

\def\normOrd#1{\mathop{:}\nolimits\!#1\!\mathop{:}\nolimits}

\begin{document}
\title{Charge transport limited by nonlocal electron--phonon interaction. I. Hierarchical equations of motion approach}

\author{Veljko Jankovi\'c}
 \email{veljko.jankovic@ipb.ac.rs}
 \affiliation{Institute of Physics Belgrade, University of Belgrade, Pregrevica 118, 11080 Belgrade, Serbia}
\begin{abstract}
Studying charge transport in models with nonlocal carrier--phonon interaction is difficult because it requires finite-temperature real-time correlation functions of mixed carrier--phonon operators.
Focusing on models with discrete undamped phonon modes, we show that such correlation functions can be retrieved from the hierarchical equations of motion (HEOM), although phonons have been integrated out.
Our procedure relies on the general explicit expression of HEOM auxiliaries in terms of phonon creation and annihilation operators.
It reveals that the auxiliaries describe multiphonon-assisted carrier transitions induced by genuine many-phonon correlations, from which lower-order correlations are subtracted according to the finite-temperature Wick's theorem.
Applying the procedure to our recently developed momentum-space HEOM method featuring appropriate hierarchy closing, we compute the numerically exact dynamical mobility of a carrier within the one-dimensional Peierls model.
The carrier mobility at moderate temperatures decreases with increasing interaction, whereas high temperatures see the opposite trend, reflecting the prevalence of the phonon-assisted current over the purely electronic band current.
The pronounced finite-size effects and HEOM instabilities delimit the range of applicability of our approach to moderate interactions, moderate to high temperatures, and not too fast phonons.
Importantly, this range comprises the values relevant for charge transport in crystalline organic semiconductors, and we present and discuss the corresponding numerically exact results in a companion paper (\href{http://arxiv.org/abs/2501.05055}{arXiv:2501.05055}).
\end{abstract}
\maketitle
\section{Introduction}
The transport of charge carriers that interact with quantum lattice vibrations has been at the forefront of both applied and fundamental research~\cite{RepProgPhys.83.036501,NatRevMater.6.560,JChemPhys.152.190902,AccChemRes.55.819,NatMater.19.491,ACSEnergyLett.6.2162,AdvMater.33.2007057}.
One of the main challenges in theoretical studies is the computationally intensive simulation of fully quantum dynamics of mutually coupled carriers and phonons~\cite{Micha-Burghardt-book}.
To reliably compute the phonon-limited carrier mobility, which is one of the key quantities in applications~\cite{AdvFunctMater.30.2001906}, such a simulation should be performed on a sufficiently large system and capture the long-time diffusive motion of the carrier.
Thus, it is not surprising that transport properties based on the fully quantum carrier--phonon dynamics remain largely inaccessible even within the simplest models of local (Holstein-type)~\cite{AnnPhys.8.325,Mahanbook,Alexandrov-Devreese-book} and nonlocal (Peierls- or Su--Schrieffer--Heeger-type)~\cite{PhysRevLett.42.1698,PhysRevB.56.4484,PhysRevLett.105.266605,PhysRevLett.96.086601} carrier--phonon interaction.

One usually contents oneself with approximate dynamics relying on physically motivated~\cite{PhysRevLett.96.086601,JChemPhys.139.174109,PhysChemChemPhys.17.12395,arXiv.2406.19851,PhysRevB.83.081202,PhysRevB.86.245201,AdvFunctMater.26.2292,PhysRevResearch.2.013001,PhysRevLett.132.266502,PhysRevB.99.104304,PhysRevX.10.021062,PhysRevB.69.075212,jcp.128.114713,PhysRevB.79.235206,PhysRevB.104.054306,PhysRevB.105.165136,AnnPhys.391.183,JPhysChemB.115.5312,mitric2024precursorsandersonlocalizationholstein} or technically convenient~\cite{PhysRevLett.103.266601,arxiv.2212.13846,PhysRevLett.91.256403,PhysRevB.74.075101,JChemTheoryComput.14.1594,JPhysChemC.122.25849,JPhysChemC.123.18804,JChemPhys.142.174103,ChemSci.12.2276,JPhysChemLett.14.3757} assumptions.
However, the domain of validity of such assumptions is \emph{a priori} unknown, and it can be determined only if some reference (numerically) exact results were available.
Recently, the method of hierarchical equations of motion (HEOM)~\cite{JPhysSocJpn.75.082001,JChemPhys.153.020901,PhysRevE.75.031107} has emerged as a reliable numerically exact method for interacting electron--phonon (or exciton--phonon) systems featuring harmonic phonons and the interaction that is linear in both phonon displacements and single-electron densities~\cite{AnnPhys.24.118}.
The HEOM method has been used to study electronic dynamics in Holstein-type models featuring a relatively small number of electronic states interacting with an \emph{infinite} number of harmonic oscillators mimicking their condensed-phase environment~\cite{ProcNatlAcadSci.106.17255,JChemTheorComput.7.2166,JChemTheoryComput.8.2808,JChemTheoryComput.11.3411,JComputChem.39.1779}.
The HEOM-based computations of transport properties of Peierls-type models in which each electronic state interacts with a \emph{finite} number of \emph{undamped} phonons~\cite{PhysRevLett.96.086601,PhysRevB.83.081202,AdvFunctMater.26.2292} face two major challenges.
(i) The dynamics exhibit numerical instabilities stemming from the discreteness of phonon spectrum~\cite{JChemPhys.150.184109,JChemPhys.153.204109,JChemPhys.160.111102}.
(ii) Apart from the purely electronic (band) contribution, the current operator, whose finite-temperature autocorrelation function determines the frequency-dependent mobility~\cite{Mahanbook,Kubo-noneq-stat-mech-book}, has a phonon-assisted contribution~\cite{PhysRev.150.529,MolPhys.18.49,PhysRevB.69.075212}.
Studying the one-dimensional Holstein model, we have resolved challenge (i) by devising an appropriate hierarchy closing scheme~\cite{JChemPhys.159.094113}.
Overcoming challenge (ii), i.e., devising the HEOM method-based framework for computing correlation functions of mixed electron--phonon operators, is the main topic of this study.

Retrieving hybrid electron--phonon dynamics from the HEOM formalism, which integrates phonons out and thus straightforwardly deals with purely electronic quantities~\cite{JChemPhys.159.094113,PhysRevB.109.214312,PhysRevB.105.054311,JChemPhys.143.194106,JChemPhys.156.244102,PhysRevLett.109.266403,JPhysChemLett.15.1382}, is a highly nontrivial task~\cite{PhysRevB.95.064308,PhysRevB.97.235429,ChemPhys.515.129}.
While it is intuitively clear that the dynamics of mixed electron--phonon quantities is related to the auxiliary operators of the HEOM formalism~\cite{JChemPhys.137.194106,JChemPhys.145.224105}, systematic connections between these ingredients had not been established before the development of the formalism of dissipaton equations of motion (DEOM)~\cite{JChemPhys.140.054105,FrontPhys.11.110306,MolPhys.116.780,JChemPhys.157.170901}.
Although the dynamical equations of the DEOM formalism are identical to those of the HEOM formalism, the former provides a physical interpretation of the auxiliary operators in terms of many-dissipaton configurations.
However, in the most general setup with dissipation, the single-dissipaton operators, the dissipaton algebra they obey, as well as their many-body configurations, remain somewhat abstract.
Moreover, the generalized Wick's theorem, which is at the crux of computing mixed electron--phonon dynamics from auxiliary operators, appears more an axiom than a theorem.  
Therefore, care should be exercised when applying the prescriptions of the DEOM formalism to compute hybrid electron--phonon dynamics in models that lack explicit dissipation~\cite{JChemPhys.150.184109,JChemPhys.153.204109,JChemPhys.160.111102}, such as the single-mode Peierls~\cite{PhysRevLett.96.086601,PhysRevB.83.081202,AdvFunctMater.26.2292} or Holstein models.

Motivated by the DEOM theory, in this study we establish a HEOM-based framework for studying carrier transport in a model with nonlocal carrier--phonon interaction and one discrete undamped phonon mode.
We explicitly express the HEOM auxiliary operators in terms of phonon creation and annihilation operators.
Our expression is quite general as it does not rely on the specific properties of the model (e.g., local or nonlocal interaction), but only on the assumptions of harmonic phonons and linear carrier--phonon interaction~\cite{AnnPhys.24.118}.
It reveals that the HEOM auxiliaries at level $n$ contain only the \emph{essential} information about $n$-phonon assisted electronic transitions, omitting the information already encoded at shallower levels.
The electronic transitions described at level $n$ are assisted by $n$-phonon correlations from which lower-order correlations are subtracted according to the prescription valid in thermal equilibrium.
Using the expression derived, we rigorously prove the generalized Wick's theorem~\cite{JChemPhys.140.054105,FrontPhys.11.110306,MolPhys.116.780,JChemPhys.157.170901}, which we subsequently use to formulate the HEOM-based framework for computing the autocorrelation function of the current operator containing both band and phonon-assisted contributions.
We discuss in detail the approximations involved in different hierarchy closing schemes, and provide solid evidence that the scheme we developed in Ref.~\onlinecite{JChemPhys.159.094113} stabilizes long-time HEOM dynamics without appreciably affecting the carrier mobility.
While the transport at moderate temperatures and interactions is dominated by the purely electronic part of the current operator, the phonon-assisted current becomes increasingly important as the temperature and/or interaction are increased.
We conclude that our framework is practically applicable only at moderate-to-high temperatures and for not excessively strong interactions.
Remarkably, it is precisely this parameter range that is relevant for carrier transport in high-mobility organic semiconductors~\cite{JChemPhys.152.190902,NatMater.19.491,PhysRevLett.96.086601,PhysRevB.83.081202,AdvFunctMater.26.2292}.
Our companion paper~\cite{part2} presents numerically exact quantum-dynamical insights into the transport of a carrier moderately coupled to slow intermolecular phonons.

The paper is structured as follows.
Section~\ref{Sec:HEOM_discrete_undamped} exposes our general results on the nature of HEOM auxiliaries and proves the generalized Wick's theorem, relegating the details to Appendices~\ref{App:HEOM_formal}--\ref{App:GWT_rigorously_proven}.
Section~\ref{Sec:HEOM_Peierls} formulates the HEOM-based computational framework for studying charge transport in the Peierls model, while Sec.~\ref{Sec:implementation_details} summarizes necessary implementation details.
In Sec.~\ref{Sec:Numerics}, we assess the applicability of the framework, and present and discuss our numerically exact results for transport properties in the field of intermediate and fast phonons.
Our main findings are summarized in Sec.~\ref{Sec:Summary}.

\section{Formal properties of hierarchical equations of motion with discrete undamped phonons}
\label{Sec:HEOM_discrete_undamped}
Having reviewed the basics of the HEOM formalism in Sec.~\ref{SSec:HEOM-basics} and Appendix~\ref{App:HEOM_formal}, we unveil the physical content of HEOM auxiliaries in models with undamped phonon modes in Sec.~\ref{SSec:many-phonon-assisted-events} and Appendix~\ref{App:Schwinger}.
In Sec.~\ref{SSec:GWT_statement} and Appendix~\ref{App:GWT_rigorously_proven}, we rigorously prove the generalized Wick's theorem~\cite{JChemPhys.140.054105,FrontPhys.11.110306,MolPhys.116.780,JChemPhys.157.170901}, which is at the heart of HEOM-based computations of real-time finite-temperature correlation functions of mixed electron--phonon operators (see Sec.~\ref{Sec:HEOM_Peierls}).

We consider a single carrier on an $N$-site chain.
We assume that each site is equipped with an undamped harmonic oscillator of frequency $\omega_0$, whose coupling $g$ to the carrier is uniform and linear in both the oscillator displacement and single-carrier densities~\cite{AnnPhys.8.325,Mahanbook,Alexandrov-Devreese-book,PhysRevB.56.4484,PhysRevLett.105.266605,PhysRevLett.96.086601}.
For definiteness, we use periodic boundary conditions, and formulate the model in momentum space.
We set the lattice constant $a_l$, the elementary charge $e_0$, and physical constants $\hbar$ and $k_B$ to unity.
The Hamiltonian reads
\begin{equation}
\label{Eq:H_tot}
\begin{split}
    H_\mathrm{tot}&=H_\mathrm{e}+H_\mathrm{ph}+H_\mathrm{e-ph}\\
    &=\sum_k\varepsilon_k |k\rangle\langle k|+\omega_0\sum_qb_q^\dagger b_q+\sum_q V_qB_q.
\end{split}
\end{equation}
The carrier ($k$) and phonon ($q$) wave numbers can assume any of the $N$ allowed values $2\pi n/N$ ($n$ is an integer) in the first Brillouin zone $(-\pi,\pi]$, and $\varepsilon_k$ is the dispersion of the free-carrier band.
The carrier--phonon interaction depends on the purely carrier operator
\begin{equation}
\label{Eq:def_V_q}
    V_q=\sum_k M(k,q)|k+q\rangle\langle k|,
\end{equation}
and the purely phononic operator
\begin{equation}
\label{Eq:def_B_q}
    B_q=\frac{g}{\sqrt{N}}(b_q+b_{\overline{q}}^\dagger).
\end{equation}
In Eq.~\eqref{Eq:def_B_q}, we define $\overline{q}=-q$, so that $B_q^\dagger=B_{\overline{q}}$ and $V_{\overline{q}}=V_q^\dagger$.
The carrier--phonon matrix element $M(k,q)$ encodes the details of the interaction (e.g., whether it is local or nonlocal).

We emphasize that the results of Secs.~\ref{SSec:HEOM-basics}--\ref{SSec:GWT_statement} are quite general as these are formulated in a manner that permits their immediate applications in coordinate space (instead of momentum space), models with nonuniform couplings [$g\to g_q$ in Eq.~\eqref{Eq:def_B_q}] or more oscillators per site. 

\subsection{Hierarchical equations of motion}
\label{SSec:HEOM-basics}

Within the HEOM formalism, the dynamics of the electronic reduced density matrix (RDM)
\begin{equation}
\label{Eq:define_rdm_t}
\begin{split}
    \rho(t)&=\mathrm{Tr}_\mathrm{ph}\left\{\rho_\mathrm{tot}(t)\right\}\\&=\mathrm{Tr}_\mathrm{ph}\left\{e^{-iH_\mathrm{tot}t}\rho_\mathrm{tot}(0)e^{iH_\mathrm{tot}t}\right\}
\end{split}
\end{equation}
is obtained by propagating the following hierarchically coupled equations~\cite{JChemPhys.153.020901,PhysRevE.75.031107,MolPhys.116.780}:
\begin{equation}
\label{Eq:HEOM-original}
\begin{split}
    &\partial_t\rho_\mathbf{n}^{(n)}(t)=-i[H_\mathrm{e},\rho_\mathbf{n}^{(n)}(t)]-\mu_\mathbf{n}\rho_\mathbf{n}^{(n)}(t)\\
    &-i\sum_{qm}\left[V_q,\rho_{\mathbf{n}_{qm}^+}^{(n+1)}(t)\right]\\
    &-i\sum_{qm}n_{qm}\sum_{q'}\left(\eta_{qq'm}V_{q'}\rho_{\mathbf{n}_{qm}^-}^{(n-1)}(t)-\eta_{\overline{q}\:\overline{q'}\:\overline{m}}^*\rho_{\mathbf{n}_{qm}^-}^{(n-1)}(t)V_{q'}\right).
\end{split}
\end{equation}
The auxiliary operator $\rho_\mathbf{n}^{(n)}(t)$ at depth $n\geq 0$ [with $\rho_\mathbf{0}^{(0)}(t)\equiv\rho(t)$] is a purely electronic operator characterized by the $2N$-dimensional vector
\begin{equation}
\label{Eq:vector_n_number_representation}
 \mathbf{n}=[n_{qm}|q;m=0,1]   
\end{equation}
of nonnegative integers $n_{qm}$ such that $n=\sum_{qm}n_{qm}$.
The auxiliaries at depth $n$ couple to auxiliaries at depths $n\pm 1$, which are characterized by vectors $\mathbf{n}_{qm}^\pm$ defined as $[\mathbf{n}_{qm}^{\pm}]_{q'm'}=n_{q'm'}\pm\delta_{q'q}\delta_{m'm}$.
The quantities $\mu_\mathbf{n}$ and $\eta_{q_2q_1m}$ are defined in Appendix~\ref{App:HEOM_formal}.

\subsection{HEOM auxiliaries and many phonon-assisted events}
\label{SSec:many-phonon-assisted-events}
The auxiliaries $\rho_\mathbf{n}^{(n)}(t)$ are most often treated as purely mathematical constructs, i.e., as intermediate quantities needed to obtain $\rho(t)$.
Physical intuition suggests that $\rho_\mathbf{n}^{(n)}(t)$ describes an $n$-phonon-assisted process whose details (quantum numbers of individual phonons and whether they are absorbed or emitted) are summarized in vector $\mathbf{n}$.
This claim is formalized by writing
\begin{equation}
\label{Eq:ADM_multiphonon_config}
    \rho_\mathbf{n}^{(n)}(t)=\mathrm{Tr}_\mathrm{ph}\left\{F_\mathbf{n}^{(n)}\rho_\mathrm{tot}(t)\right\}.
\end{equation}
The purely phononic operator $F_\mathbf{n}^{(n)}$ is to be expressed in terms of phonon creation and annihilation operators appearing in vector $\mathbf{n}$ considered as a set of $n$ pairs
\begin{equation}
\label{Eq:vector_n_momentum_representation}
    \mathbf{n}=\{(q_i,m_i)|i=1,\dots,n\}.
\end{equation} 
The order of pairs is immaterial, and some of them can be mutually equal.
Introducing operators
\begin{equation}
\label{Eq:def_f_qm}
    f_{q0}=\frac{g}{\sqrt{N}}b_q,\quad f_{q1}=\frac{g}{\sqrt{N}} b_{\overline{q}}^\dagger
\end{equation}
such that $B_q=\sum_m f_{qm}$, and abbreviating $f_{q_im_i}\equiv f_i$, in Appendix~\ref{App:Schwinger} we prove that
\begin{equation}
\label{Eq:F_n_normal_ordered}
\begin{split}
    &F_\mathbf{n}^{(n)}=\normOrd{\prod_{a=1}^n f_{a}}-\sideset{}{^n}\sum_{(ij)}\langle\normOrd{f_{j}f_{i}}\rangle_\mathrm{ph}\normOrd{\prod_{\substack{a=1\\a\neq i,j}}^n f_{a}}\\
    &+\sideset{}{^n}\sum_{(ij)(rs)}\langle\normOrd{f_{s}f_{r}}\rangle_\mathrm{ph}\langle\normOrd{f_{j}f_{i}}\rangle_\mathrm{ph}\normOrd{\prod_{\substack{a=1\\a\neq i,j,r,s}}^n f_{a}}-\dots
\end{split}
\end{equation}
In Eq.~\eqref{Eq:F_n_normal_ordered}, the normal-ordering symbol $\normOrd{\:}$ rearranges the product of $f$ operators so that the creation operators $f_{q1}$ are to the left of all annihilation operators $f_{q0}$, while $\langle O_\mathrm{ph}\rangle_\mathrm{ph}=\mathrm{Tr}_\mathrm{ph}\{O_\mathrm{ph}\rho_\mathrm{ph}^\mathrm{eq}\}$ denotes the average of a purely phononic operator $O_\mathrm{ph}$ in the phonon equilibrium $\rho_\mathrm{ph}^\mathrm{eq}=\frac{e^{-\beta H_\mathrm{ph}}}{\mathrm{Tr}_\mathrm{ph}\:e^{-\beta H_\mathrm{ph}}}$ at temperature $T=\beta^{-1}$.
In the second term on the RHS of Eq.~\eqref{Eq:F_n_normal_ordered}, the sum $\displaystyle{\sideset{}{^n}\sum_{(ij)}}$ runs over $\binom{n}{2}$ pairs $(ij)$ that can be chosen out of $n$ elements $\{1,\dots,n\}$.
The sum $\displaystyle{\sideset{}{^n}\sum_{(ij)(rs)}}$ in the third term on the RHS of Eq.~\eqref{Eq:F_n_normal_ordered} runs over $\frac{1}{2}\binom{n}{2}\binom{n-2}{2}$ double pairs $(ij)(rs)$ that can be chosen from $\{1,\dots,n\}$.

Apart from the normally ordered product of $n$ phonon operators, Eq.~\eqref{Eq:F_n_normal_ordered} also contains normally ordered products of $n-2$, $n-4$, etc. phonon operators appearing with alternating signs.
This form somewhat resembles the cluster-expansion approach to quantum dynamics~\cite{AnnPhys.159.328,ProgQuantumElectron.30.155,PhysRevA.73.013813,Kirabook}, and suggests that $F_\mathbf{n}^{(n)}$ describes genuine $n$-phonon correlations, from which lower-order many-phonon correlations are subtracted.
Indeed, in Appendix~\ref{App:Schwinger} we also derive the following expression for $F_\mathbf{n}^{(n)}$, in which the subtraction of lower-order $F$ operators from the normally ordered product of $n$ phonon operators is manifest:
\begin{equation}
\label{Eq:F_n_F_n}
\begin{split}
    &F_\mathbf{n}^{(n)}=\normOrd{\prod_{a=1}^n f_{a}}-\sideset{}{^n}\sum_{(ij)}\langle\normOrd{f_{j}f_{i}}\rangle_\mathrm{ph}F_{\mathbf{n}_{ji}^-}^{(n-2)}\\
    &-\sideset{}{^n}\sum_{(ij)(rs)}\langle\normOrd{f_{s}f_{r}}\rangle_\mathrm{ph}\langle\normOrd{f_{j}f_{i}}\rangle_\mathrm{ph}F_{\mathbf{n}_{srji}^-}^{(n-4)}-\dots
\end{split}
\end{equation}
In Eq.~\eqref{Eq:F_n_F_n}, $\mathbf{n}_{ji}^-=\mathbf{n}\setminus\{(q_j,m_j),(q_i,m_i)\}$.

The Wick's theorem at finite temperature shows that $\left\langle F_\mathbf{n}^{(n)}\right\rangle_\mathrm{ph}=\delta_{n,0}$, i.e., the choice of $F_\mathbf{n}^{(n)}$ embodied in Eqs.~\eqref{Eq:F_n_normal_ordered} or~\eqref{Eq:F_n_F_n} provides the most convenient representation of many-phonon correlations in thermal equilibrium.
In the time-dependent setup [Eq.~\eqref{Eq:ADM_multiphonon_config}], Eq.~\eqref{Eq:F_n_F_n} suggests that the HEOM auxiliaries at level $n$ remove lower-order many-phonon correlations only partially, effectively assuming that phonons are in thermal equilibrium.
This assumption is often used when studying the coupled carrier--phonon dynamics in, e.g., photoexcited semiconductors~\cite{PhysRevB.50.5435,RevModPhys.70.145,PhysRevB.92.235208}.
On the other hand, cluster expansion-based approaches~\cite{AnnPhys.159.328,ProgQuantumElectron.30.155,PhysRevA.73.013813,Kirabook} strive to fully remove the dynamical lower-order many-phonon correlations by using the time-dependent expectation values instead of the equilibrium expectation values entering Eq.~\eqref{Eq:F_n_F_n}.

\subsection{Generalized Wick's theorem}
\label{SSec:GWT_statement}
Using Eq.~\eqref{Eq:F_n_normal_ordered}, in Appendix~\ref{App:GWT_rigorously_proven} we prove the so-called generalized Wick's theorem
\begin{equation}
\label{Eq:GWT_right}
\begin{split}
    F_{\mathbf{n}}^{(n)}f_{n+1}=&F_{\mathbf{n}_{n+1}^+}^{(n+1)}+\sum_{i=1}^n\langle f_{i}f_{n+1}\rangle_\mathrm{ph}F_{\mathbf{n}_{i}^-}^{(n-1)},
\end{split}
\end{equation}
\begin{equation}
\label{Eq:GWT_left}
\begin{split}
    f_{n+1}F_{\mathbf{n}}^{(n)}=&F_{\mathbf{n}_{n+1}^+}^{(n+1)}+\sum_{i=1}^n\langle f_{n+1}f_{i}\rangle_\mathrm{ph}F_{\mathbf{n}_{i}^-}^{(n-1)}.
\end{split}
\end{equation}
In Eqs.~\eqref{Eq:GWT_right} and~\eqref{Eq:GWT_left}, $\mathbf{n}$ is defined as in Eq.~\eqref{Eq:vector_n_momentum_representation}, and $\mathbf{n}_{n+1}^+=\mathbf{n}\cup\{(q_{n+1},m_{n+1})\}$.
If we put emphasis on the number of $n_{qm}$ of phonon-assisted events with momentum $q$ and type $m$ and use the definition of $\mathbf{n}$ in Eq.~\eqref{Eq:vector_n_number_representation}, as well as Eq.~\eqref{Eq:2-point-expt-values}, we rewrite the generalized Wick's theorem in the form in which it appears in DEOM references~\cite{JChemPhys.140.054105,FrontPhys.11.110306,MolPhys.116.780,JChemPhys.157.170901}
\begin{equation}
  \label{Eq:GWT_right_DEOM}
  F_{\mathbf{n}}^{(n)}f_{qm}=F_{\mathbf{n}_{qm}^+}^{(n+1)}+\sum_{q'm'}n_{q'm'}
  \eta_{q'qm'}F_{\mathbf{n}_{q'm'}^-}^{(n-1)},
\end{equation}
\begin{equation}
\label{Eq:GWT_left_DEOM}
  f_{qm}F_{\mathbf{n}}^{(n)}=F_{\mathbf{n}_{qm}^+}^{(n+1)}+\sum_{q'm'}n_{q'm'}
  \eta_{\overline{q'}\:\overline{q}\:\overline{m'}}^*F_{\mathbf{n}_{q'm'}^-}^{(n-1)}.
\end{equation}

In Sec.~SI of the Supplemental Material~\cite{comment241224}, we discuss how the generalized Wick's theorem can be inferred from the dynamical equations of the HEOM formalism [Eq.~\eqref{Eq:HEOM-original}] themselves.

\section{HEOM-based theory of charge transport in the Peierls model}
\label{Sec:HEOM_Peierls}
Exploiting the formal results of Sec.~\ref{Sec:HEOM_discrete_undamped}, we formulate a HEOM-based framework for studying carrier transport in a widely studied model~\cite{PhysRevB.56.4484,PhysRevLett.96.086601,PhysRevLett.105.266605,PhysRevLett.114.086601,AdvFunctMater.26.2292} with nonlocal carrier--phonon interaction.
In the limit of low carrier density, transport dynamics are encoded in the real-time current--current correlation function~\cite{JChemPhys.159.094113,PhysRevB.109.214312}
\begin{equation}
\label{Eq:def_C_jj}
\begin{split}
    C_{jj}(t)=\langle j(t)j(0)\rangle=\mathrm{Tr}\left\{je^{-iH_\mathrm{tot}t}j\rho_\mathrm{tot}^\mathrm{eq}e^{iH_\mathrm{tot}t}\right\}.
\end{split}
\end{equation}
The angular brackets $\langle\cdot\rangle$ in Eq.~\eqref{Eq:def_C_jj} denote averaging over the equilibrium state
\begin{equation}
\label{Eq:interacting_eq}
    \rho_\mathrm{tot}^\mathrm{eq}=\frac{e^{-\beta H_\mathrm{tot}}}{\mathrm{Tr}\left\{e^{-\beta H_\mathrm{tot}}\right\}}
\end{equation}
of the interacting electron--phonon system.

In the one-dimensional Peierls model, the nearest-neighbor hopping amplitude $J$ [giving rise to the free-carrier dispersion $\varepsilon_k=-2J\cos k$ in Eq.~\eqref{Eq:H_tot}] is modulated by the difference between coordinates of the corresponding local oscillators~\cite{PhysRevB.56.4484,PhysRevLett.96.086601,PhysRevLett.105.266605,PhysRevLett.114.086601,AdvFunctMater.26.2292}.
The carrier--phonon matrix element $M(k,q)$ [Eq.~\eqref{Eq:def_V_q}] then reads~\cite{JChemPhys.100.2335}
\begin{equation}
\label{Eq:def_M_k_q}
    M(k,q)=-2i\left[\sin(k+q)-\sin k\right].
\end{equation}
Equation~\eqref{Eq:def_M_k_q} implies that the totally symmetric phonon mode ($q=0$) is exactly uncoupled from the remaining phonon modes and carrier states.
In the following, it is understood that the $q=0$ term is excluded from all summations over phonon wave number $q$, as well as from vector $\mathbf{n}$ in Eq.~\eqref{Eq:vector_n_number_representation}, which contains $2(N-1)$ nonnegative integers $n_{qm}$ (for $q\neq 0$ and $m=0,1$).

The current operator is
\begin{equation}
\label{Eq:def_j_tot}
    j=j_\mathrm{e}+j_\mathrm{e-ph},
\end{equation}
where the purely electronic contribution
\begin{equation}
\label{Eq:def_j_e}
    j_\mathrm{e}=\sum_k v_kP_k
\end{equation}
describes the band conduction, while the phonon-assisted contribution is
\begin{equation}
\label{Eq:def_j_e-ph}
    j_\mathrm{e-ph}=\sum_q J_qB_q=\sum_{qm}J_qf_{qm}.
\end{equation}
In Eqs.~\eqref{Eq:def_j_e} and~\eqref{Eq:def_j_e-ph}, $v_k=\frac{\partial\varepsilon_k}{\partial k}$ is the band velocity, $P_k=|k\rangle\langle k|$, while
\begin{equation}
\label{Eq:def_J_q}
    J_q=\sum_k M_J(k,q)|k+q\rangle\langle k|,
\end{equation}
with
\begin{equation}
\label{Eq:def_M_J_k_q}
    M_J(k,q)=\frac{\partial M(k,q)}{\partial k}=-2i\left[\cos(k+q)-\cos k\right],
\end{equation}
is a purely electronic operator increasing the electronic momentum by $q$ and satisfying $J_q=J_{\overline{q}}^\dagger$.

The central object of our HEOM-based framework is the operator
\begin{equation}
\label{Eq:def_iota_tot_t}
\begin{split}
    \iota_\mathrm{tot}(t)=e^{-iH_\mathrm{tot}t}j\rho_\mathrm{tot}^\mathrm{eq}e^{iH_\mathrm{tot}t}
\end{split}
\end{equation}
in terms of which Eq.~\eqref{Eq:def_C_jj} reads $C_{jj}(t)=\mathrm{Tr}\left\{j\iota_\mathrm{tot}(t)\right\}$.
Although Sec.~\ref{SSec:many-phonon-assisted-events} deals with the RDM $\rho(t)$ and the corresponding auxiliaries $\rho_\mathbf{n}^{(n)}(t)$, its results are quite general as these do not rely on the properties of the density matrix (hermiticity, normalization), but only on the properties of phonons (Gaussian statistics, finite-temperature Wick's theorem) and the electron--phonon interaction (linear in phonon displacements and electronic densities)~\cite{PhysRep.168.115}.
Even though the operator $\iota_\mathrm{tot}(t)$ is nonhermitean, it is determined by the HEOM embodied in Eq.~\eqref{Eq:HEOM-original} for the auxiliaries defined by [see Eqs.~\eqref{Eq:ADM_multiphonon_config} and~\eqref{Eq:F_n_normal_ordered}]
\begin{equation}
\label{Eq:define_iota_mathbf-n_n}
    \iota_\mathbf{n}^{(n)}(t)=\mathrm{Tr}_\mathrm{ph}\{F_\mathbf{n}^{(n)}\iota_\mathrm{tot}(t)\}.
\end{equation}
The corresponding initial conditions [see Eqs.~\eqref{Eq:def_iota_tot_t} and~\eqref{Eq:def_j_tot}]
\begin{equation}
\label{Eq:def_iota_aux_t_0}
\begin{split}
    &\iota_\mathbf{n}^{(n)}(0)\equiv\iota_\mathbf{n}^{(n,\mathrm{eq})}=\iota_{\mathrm{e},\mathbf{n}}^{(n,\mathrm{eq})}+\iota_{\mathrm{e-ph},\mathbf{n}}^{(n,\mathrm{eq})}\\
    &=\mathrm{Tr}_\mathrm{ph}\left\{F_\mathbf{n}^{(n)}j_\mathrm{e}\rho_\mathrm{tot}^\mathrm{eq}\right\}+\mathrm{Tr}_\mathrm{ph}\left\{F_\mathbf{n}^{(n)}j_\mathrm{e-ph}\rho_\mathrm{tot}^\mathrm{eq}\right\}
\end{split}
\end{equation}
are fixed by the HEOM representation $\{\rho_\mathbf{n}^{(n,\mathrm{eq})}=\mathrm{Tr}_\mathrm{ph}\{F_\mathbf{n}^{(n)}\rho_\mathrm{tot}^\mathrm{eq}\}\}$ of $\rho_\mathrm{tot}^\mathrm{eq}$ [Eq.~\eqref{Eq:interacting_eq}], which is discussed below.
The contribution to Eq.~\eqref{Eq:def_iota_aux_t_0} that depends on the band current is
\begin{equation}
\label{Eq:iota_e_n_eq}
    \iota_{\mathrm{e},\mathbf{n}}^{(n,\mathrm{eq})}=\mathrm{Tr}_\mathrm{ph}\left\{F_\mathbf{n}^{(n)}j_\mathrm{e}\rho_\mathrm{tot}^\mathrm{eq}\right\}=\sum_k v_kP_k\rho_\mathbf{n}^{(n,\mathrm{eq})}.
\end{equation}
The contribution containing the phonon-assisted current is evaluated using the generalized Wick's theorem [Eq.~\eqref{Eq:GWT_right_DEOM}]
\begin{equation}
\label{Eq:iota_e-ph_n_eq}
\begin{split}
    &\iota_{\mathrm{e-ph},\mathbf{n}}^{(n,\mathrm{eq})}=\mathrm{Tr}_\mathrm{ph}\left\{F_\mathbf{n}^{(n)}j_\mathrm{e-ph}\rho_\mathrm{tot}^\mathrm{eq}\right\}=\\&\sum_{qm}\left(J_q\rho_{\mathbf{n}_{qm}^+}^{(n+1,\mathrm{eq})}+n_{qm}\sum_{q'}\eta_{qq'm}J_{q'}\rho_{\mathbf{n}_{qm}^-}^{(n-1,\mathrm{eq})}\right).
\end{split}
\end{equation}
One then propagates the real-time HEOM for $\iota_\mathbf{n}^{(n)}(t)=\iota_{\mathrm{e},\mathbf{n}}^{(n)}(t)+\iota_{\mathrm{e-ph},\mathbf{n}}^{(n)}(t)$, see Eq.~\eqref{Eq:HEOM-original}, with the initial conditions in Eq.~\eqref{Eq:def_iota_aux_t_0}.
At each instant $t$, one inserts Eqs.~\eqref{Eq:def_j_tot}--\eqref{Eq:def_j_e-ph} and~\eqref{Eq:def_iota_tot_t} into Eq.~\eqref{Eq:def_C_jj}, and uses $F_\mathbf{0}^{(0)}=\mathbbm{1}_\mathrm{ph}$ and $F_{\mathbf{0}_{qm}^+}^{(1)}=f_{qm}$ [see Eq.~\eqref{Eq:F_n_normal_ordered}] to finally obtain
\begin{equation}
\label{Eq:C_jj_t_final}
\begin{split}
    &C_{jj}(t)=\mathrm{Tr}\{j\iota_\mathrm{tot}(t)\}=\\&\sum_k v_k\mathrm{Tr}_\mathrm{e}\left\{P_k\iota_\mathbf{0}^{(0)}(t)\right\}+\sum_{qm}\mathrm{Tr}_\mathrm{e}\left\{J_q\iota_{\mathbf{0}_{qm}^+}^{(1)}(t)\right\}.
\end{split}
\end{equation}

Equations~\eqref{Eq:iota_e-ph_n_eq} and~\eqref{Eq:C_jj_t_final} overcome the long-standing issue with the phonon-assisted current within the HEOM formalism for discrete undamped phonons.
From a broader perspective, these equations show the utility of the generalized Wick's theorem in the computation of real-time finite-temperature correlation functions of mixed electron--phonon operators.
Within the model considered here, these equations enable us to separately analyze different contributions to the current--current correlation function and gain important physical insights into the character of charge transport.
The decomposition of the current operator in Eq.~\eqref{Eq:def_j_tot} implies that $C_{jj}(t)$ can be decomposed as
\begin{equation}
\label{Eq:decomposition_C_jj}
    C_{jj}(t)=C_\mathrm{e}(t)+C_\mathrm{ph}(t)+C_\mathrm{x}(t),
\end{equation}
where
\begin{equation}
\label{Eq:def_C_e}
    C_\mathrm{e}(t)=\langle j_\mathrm{e}(t)j_\mathrm{e}(0)\rangle=\sum_k v_k\mathrm{Tr}_\mathrm{e}\left\{P_k\iota_{\mathrm{e},\mathbf{0}}^{(0)}(t)\right\}
\end{equation}
is the purely electronic (band) contribution,
\begin{equation}
\label{Eq:def_C_ph}
\begin{split}
    C_\mathrm{ph}(t)&=\langle j_\mathrm{e-ph}(t)j_\mathrm{e-ph}(0)\rangle\\
    &=\sum_{qm}\mathrm{Tr}_\mathrm{e}\left\{J_q\iota_{\mathrm{e-ph},{\mathbf{0}_{qm}^+}}^{(1)}(t)\right\}
\end{split}
\end{equation}
is the phonon-assisted contribution, while
\begin{equation}
\label{Eq:def_C_x}
\begin{split}
    &C_\mathrm{x}(t)=\langle j_\mathrm{e}(t)j_\mathrm{e-ph}(0)\rangle+\langle j_\mathrm{e-ph}(t)j_\mathrm{e}(0)\rangle\\
    &=\sum_k v_k\mathrm{Tr}_\mathrm{e}\left\{P_k\iota_{\mathrm{e-ph},\mathbf{0}}^{(0)}(t)\right\}+\sum_{qm}\mathrm{Tr}_\mathrm{e}\left\{J_q\iota_{\mathrm{e},{\mathbf{0}_{qm}^+}}^{(1)}(t)\right\}
\end{split}
\end{equation}
is the cross contribution to $C_{jj}$.

The HEOM representation of $\rho_\mathrm{tot}^\mathrm{eq}$ is obtained by propagating the following imaginary-time HEOM
\begin{equation}
\label{Eq:im-time-HEOM-original}
\begin{split}
    &\partial_\tau\sigma_\mathbf{n}^{(n)}(\tau)=-H_\mathrm{e}\sigma_\mathbf{n}^{(n)}(\tau)+i\mu_\mathbf{n}\sigma_\mathbf{n}^{(n)}(\tau)\\
    &-\sum_{qm}V_q\sigma_{\mathbf{n}_{qm}^+}^{(n+1)}(\tau)\\
    &-\sum_{qm}n_{qm}\sum_{q'}\eta_{qq'm}V_{q'}\sigma_{\mathbf{n}_{qm}^-}^{(n-1)}(\tau)
\end{split}
\end{equation}
from $\tau=0$ to $\beta$ with the infinite-temperature initial condition $\sigma_\mathbf{n}^{(n)}(0)=\delta_{n,0}\mathbbm{1}_\mathrm{e}$~\cite{PhysRevB.105.054311,JChemPhys.159.094113}.
The operator $\rho_\mathbf{n}^{(n,\mathrm{eq})}$ then reads~\cite{PhysRevB.105.054311,JChemPhys.159.094113}
\begin{equation}
\label{Eq:heom_representation_eq_tot}
    \rho_\mathbf{n}^{(n,\mathrm{eq})}=\frac{\sigma_\mathbf{n}^{(n)}(\beta)}{\sum_k\langle k|\sigma_\mathbf{0}^{(0)}(\beta)|k\rangle}.
\end{equation}

\section{Numerical implementation}
\label{Sec:implementation_details}
\subsection{Rescaled and dimensionless momentum-space HEOM}
The momentum conservation implies that the only non-zero matrix elements of real-time HEOM auxiliaries $\iota_\mathbf{n}^{(n)}(t)$ [Eq.~\eqref{Eq:define_iota_mathbf-n_n}] are $\langle k|\iota_\mathbf{n}^{(n)}(t)|k+k_\mathbf{n}\rangle$, and similarly for imaginary-time HEOM auxiliaries $\sigma_\mathbf{n}^{(n)}(\tau)$ [Eq.~\eqref{Eq:im-time-HEOM-original}]~\cite{PhysRevB.105.054311,JChemPhys.159.094113,PhysRevB.109.214312}.
Here,
\begin{equation}
    k_\mathbf{n}=\sum_{qm}qn_{qm}    
\end{equation}
is the total momentum exchanged between the carrier and phonons in the multiphonon-assisted process described by vector $\mathbf{n}$.
Instead of $\sigma_\mathbf{n}^{(n)}(\tau)$ and $\iota_\mathbf{n}^{(n)}(t)$, our numerical implementation considers the following rescaled and dimensionless auxiliaries: 
\begin{equation}
    \widetilde{\sigma}_\mathbf{n}^{(n)}(\tau)=f(\mathbf{n})\sigma_\mathbf{n}^{(n)}(\tau),\quad\widetilde{\iota}_\mathbf{n}^{(n)}(t)=f(\mathbf{n})\iota_\mathbf{n}^{(n)}(t),
\end{equation}
where the rescaling factor $f(\mathbf{n})$ reads [with $c_m$ defined in Eqs.~\eqref{Eq:def_c_0} and~\eqref{Eq:def_c_1}]~\cite{JChemPhys.130.084105}
\begin{equation}
    f(\mathbf{n})=\prod_{qm}\left(|c_m|^{n_{qm}}n_{qm}!\right)^{-1/2}.
\end{equation}

The rescaled and dimensionless imaginary-time momentum-space HEOM reads [see Eq.~\eqref{Eq:im-time-HEOM-original}]
\begin{equation}
\label{Eq:im-time-HEOM-final}
\begin{split}
&\partial_\tau\langle k|\widetilde{\sigma}_\mathbf{n}^{(n)}(\tau)|k+k_\mathbf{n}\rangle=-\left(\varepsilon_k-i\mu_\mathbf{n}\right)\langle k|\widetilde{\sigma}_\mathbf{n}^{(n)}(\tau)|k+k_\mathbf{n}\rangle\\
&-\sum_{qm}\sqrt{1+n_{qm}}\sqrt{|c_m|}\:M(k-q,q)\langle k-q|\widetilde{\sigma}_{\mathbf{n}_{qm}^+}^{(n+1)}(\tau)|k+k_\mathbf{n}\rangle\\
&-\sum_{qm}\sqrt{n_{qm}}\sqrt{|c_m|}\:M(k+q,-q)\langle k+q|\widetilde{\sigma}_{\mathbf{n}_{qm}^-}^{(n-1)}(\tau)|k+k_\mathbf{n}\rangle.
\end{split}
\end{equation}
The rescaled and dimensionless HEOM representation $\{\widetilde{\rho}_\mathbf{n}^{(n,\mathrm{eq})}\}$ of the equilibrium state $\rho_\mathrm{tot}^\mathrm{eq}$ of the interacting electron--phonon system is then given by Eq.~\eqref{Eq:heom_representation_eq_tot}.
The matrix elements of the rescaled and dimensionless HEOM representation $\{\widetilde{\iota}_\mathbf{n}^{(n,\mathrm{eq})}=\widetilde{\iota}_{\mathrm{e},\mathbf{n}}^{(n,\mathrm{eq})}+\widetilde{\iota}_{\mathrm{e-ph},\mathbf{n}}^{(n,\mathrm{eq})}\}$ of the operator $j\rho_\mathrm{tot}^\mathrm{eq}$ are [see Eqs.~\eqref{Eq:iota_e_n_eq} and~\eqref{Eq:iota_e-ph_n_eq}]
\begin{equation}
    \langle k|\widetilde{\iota}_{\mathrm{e},\mathbf{n}}^{(n,\mathrm{eq})}|k+k_\mathbf{n}\rangle=v_k\langle k|\widetilde{\rho}_{\mathrm{e},\mathbf{n}}^{(n,\mathrm{eq})}|k+k_\mathbf{n}\rangle,
\end{equation}
\begin{equation}
\begin{split}
    &\langle k|\widetilde{\iota}_{\mathrm{e-ph},\mathbf{n}}^{(n,\mathrm{eq})}|k+k_\mathbf{n}\rangle=\\&\sum_{qm}\sqrt{1+n_{qm}}\sqrt{|c_m|}\:M_J(k-q,q)\langle k-q|\widetilde{\rho}_{\mathbf{n}_{qm}^+}^{(n+1,\mathrm{eq})}|k+k_\mathbf{n}\rangle\\
    &+\sum_{qm}\sqrt{n_{qm}}\frac{c_m}{\sqrt{|c_m|}}\:M_J(k+q,-q)\langle k+q|\widetilde{\rho}_{\mathbf{n}_{qm}^-}^{(n-1,\mathrm{eq})}|k+k_\mathbf{n}\rangle.
\end{split}
\end{equation}

The rescaled and dimensionless real-time momentum-space HEOM for $\widetilde{\iota}_\mathbf{n}^{(n)}(t)$ reads [see Eq.~\eqref{Eq:HEOM-original}]
\begin{widetext}
\begin{equation}
\label{Eq:re-time-HEOM-final}
\begin{split}
\partial_t\langle k|\widetilde{\iota}_\mathbf{n}^{(n)}(t)|k+k_\mathbf{n}\rangle=&-i\left(\varepsilon_k-\varepsilon_{k+k_\mathbf{n}}-i\mu_\mathbf{n}\right)\langle k|\widetilde{\iota}_\mathbf{n}^{(n)}(t)|k+k_\mathbf{n}\rangle\\
&-i\sum_{qm}\sqrt{1+n_{qm}}\sqrt{|c_m|}\:M(k-q,q)\langle k-q|\widetilde{\iota}_{\mathbf{n}_{qm}^+}^{(n+1)}(t)|k+k_\mathbf{n}\rangle\\
&+i\sum_{qm}\sqrt{1+n_{qm}}\sqrt{|c_m|}\:M(k+k_\mathbf{n},q)\langle k|\widetilde{\iota}_{\mathbf{n}_{qm}^+}^{(n+1)}(t)|k+k_\mathbf{n}+q\rangle\\
&-i\sum_{qm}\sqrt{n_{qm}}\frac{c_m}{\sqrt{|c_m|}}\:M(k+q,-q)\langle k+q|\widetilde{\iota}_{\mathbf{n}_{qm}^-}^{(n-1)}(t)|k+k_\mathbf{n}\rangle\\
&+i\sum_{qm}\sqrt{n_{qm}}\frac{c_{\overline{m}}^*}{\sqrt{|c_m|}}\:M(k+k_\mathbf{n},-q)\langle k|\widetilde{\iota}_{\mathbf{n}_{qm}^-}^{(n-1)}(t)|k+k_\mathbf{n}-q\rangle\\
&+[\partial_t\langle k|\widetilde{\iota}_\mathbf{n}^{(n)}(t)|k+k_\mathbf{n}\rangle]_\mathrm{close}.
\end{split}
\end{equation}
\end{widetext}
The same equations govern the dynamics of its contributions $\widetilde{\iota}_{\mathrm{e},\mathbf{n}}^{(n)}(t)$ and $\widetilde{\iota}_{\mathrm{e-ph},\mathbf{n}}^{(n)}(t)$.
The closing term $[\partial_t\langle k|\widetilde{\iota}_\mathbf{n}^{(n)}(t)|k+k_\mathbf{n}\rangle]_\mathrm{close}$ is discussed in Sec.~\ref{SSec:heom_closing_schemes}.

Finally, different contributions to the current--current correlation functions are computed as [see Eqs.~\eqref{Eq:def_C_e}--\eqref{Eq:def_C_x}]
\begin{equation}
\label{Eq:C_e_t_final}
    C_\mathrm{e}(t)=\sum_k v_k\langle k|\widetilde{\iota}_{\mathrm{e},\mathbf{0}}^{(0)}(t)|k\rangle,
\end{equation}
\begin{equation}
\label{Eq:C_ph_t_final}
    C_\mathrm{ph}(t)=\sum_{qmk}\sqrt{|c_{m}|}M_J(k,q)\langle k|\widetilde{\iota}_{\mathrm{e-ph},\mathbf{0}_{qm}^+}^{(1)}(t)|k+q\rangle,
\end{equation}
\begin{equation}
\label{Eq:C_x_t_final}
\begin{split}
    &C_\mathrm{x}(t)=\sum_k v_k\langle k|\widetilde{\iota}_{\mathrm{e-ph},\mathbf{0}}^{(0)}(t)|k\rangle+\\&\sum_{qmk}\sqrt{|c_{m}|}M_J(k,q)\langle k|\widetilde{\iota}_{\mathrm{e},\mathbf{0}_{qm}^+}^{(1)}(t)|k+q\rangle.
\end{split}
\end{equation}

\subsection{HEOM closing schemes}
\label{SSec:heom_closing_schemes}
In actual computations, we truncate both the imaginary-time HEOM in Eq.~\eqref{Eq:im-time-HEOM-final} and the real-time HEOM in Eq.~\eqref{Eq:re-time-HEOM-final} at the same maximum depth $D$.
In models with discrete undamped phonons, the truncated real-time HEOM is known to exhibit numerical instabilities at sufficiently long times~\cite{JChemPhys.150.184109,JChemPhys.153.204109,JChemPhys.160.111102}, which severely hamper accurate computations of carrier mobility.
The instabilities can be overcome by devising a hierarchy closing scheme~\cite{JChemPhys.159.094113}, which amounts to approximately solving the dynamical equations at depth $D+1$ in terms of the auxiliaries at depth $D$, see Appendix~\ref{App:hierarchy_closing}.
We simplify the resulting closing term, which introduces couplings between auxiliaries at depth $D$, by invoking the random-phase approximation~\cite{kuhncontribution,RevModPhys.74.895,PhysRevB.92.235208}, which neglects momentum-averaged matrix elements of auxiliaries at depth $D$ due to random phases at different momenta.
Therefore, our general closing term reads
\begin{equation}
\label{Eq:closing_most_general}
\begin{split}
    &[\partial_t\langle k|\widetilde{\iota}_\mathbf{n}^{(n)}(t)|k+k_\mathbf{n}\rangle]_\mathrm{close}=\\
    &-\delta_{n,D}\Gamma(k,\mathbf{n})\langle k|\widetilde{\iota}_\mathbf{n}^{(n)}(t)|k+k_\mathbf{n}\rangle.
\end{split}
\end{equation}

In Appendix~\ref{SApp:MA_closing}, we solve the equations at depth $D+1$ in the Markovian and adiabatic (MA) approximations and obtain the MA closing term~\cite{JChemPhys.159.094113}
\begin{equation}
\label{Eq:MA_closing}
\begin{split}
    \Gamma_\mathrm{MA}(k,\mathbf{n})=\frac{1}{2}\left(\tau_k^{-1}+\tau_{k+k_\mathbf{n}}^{-1}\right).
\end{split}
\end{equation}
In Eq.~\eqref{Eq:MA_closing}, $\tau_k^{-1}$ is the quasiparticle scattering rate out of the free-electron state $|k\rangle$ computed in the second-order perturbation theory and in the infinite-chain limit.
The corresponding expression reads (see, e.g., Ref.~\onlinecite{PhysRevResearch.2.013001})
\begin{equation}
\label{Eq:momentum_relaxation_time}
\begin{split}
    &\tau_k^{-1}=\frac{4g^2}{J\left(e^{\beta\omega_0}-1\right)}\frac{2-\left(\frac{\varepsilon_k}{2J}\right)^2-\left(\frac{\varepsilon_k+\omega_0}{2J}\right)^2}{\sqrt{1-\left(\frac{\varepsilon_k+\omega_0}{2J}\right)^2}}\\&+\frac{4g^2}{J\left(1-e^{-\beta\omega_0}\right)}\frac{2-\left(\frac{\varepsilon_k}{2J}\right)^2-\left(\frac{\varepsilon_k-\omega_0}{2J}\right)^2}{\sqrt{1-\left(\frac{\varepsilon_k-\omega_0}{2J}\right)^2}}.
\end{split}
\end{equation}
It is known that, for $\omega_0/J\geq 2$, there exist $k$-states such that $\tau_k^{-1}=0$~\cite{PhysRevB.99.104304,arxiv.2212.13846}, which underlies the ineffectiveness of our closing scheme in the antiadiabatic regime~\cite{JChemPhys.159.094113}.
An \emph{ad hoc} solution to this problem is to replace Eq.~\eqref{Eq:MA_closing} with
\begin{equation}
\label{Eq:prescription-antiadiabatic}
\begin{split}
    \Gamma_\mathrm{MA-avg}(k,\mathbf{n})=\frac{1}{N}\sum_p\tau_p^{-1}.
\end{split}
\end{equation}
Equation~\eqref{Eq:prescription-antiadiabatic} is used to obtain the results for $\omega_0/J=3$ in Figs.~\ref{Fig:mu_vs_T_omega_3_overall_241024} and~\ref{Fig:details_rearranged_071024}(d).

The hierarchy closing scheme represents the main approximation of our framework, which otherwise provides an \emph{exact} treatment of phonon-assisted processes up to order $D$.
For sufficiently large $D$, we expect that the dynamics of the zeroth- and first-tier auxiliaries determining $C_{jj}(t)$ [see Eqs.~\eqref{Eq:decomposition_C_jj} and~\eqref{Eq:C_e_t_final}--\eqref{Eq:C_x_t_final}] is stabilized in a manner that weakly depends on the particular form of the closing terms $\Gamma(k,\mathbf{n})$ entering Eq.~\eqref{Eq:closing_most_general}.
To confirm this expectation, in Appendix~\ref{SApp:DR_closing} we generalize the derivative-resum (DR) closing scheme originally developed in Refs.~\cite{JChemPhys.142.104112,JChemPhys.157.054108} to our undamped-phonon model.
We compare and contrast representative examples employing MA and DR schemes in Sec.~\ref{SSec:influence_heom_closing_on_converged}. 

\subsection{Further technical details}
\label{SSec:technical-details}
Both the real-time HEOM [Eq.~\eqref{Eq:re-time-HEOM-final}] and the imaginary-time HEOM [Eq.~\eqref{Eq:im-time-HEOM-final}] are propagated with the timestep $J\Delta t=J\Delta\tau=(1-2)\times 10^{-2}$ using the propagation scheme from Ref.~\onlinecite{JChemTheoryComput.11.3411}.
We propagate the real-time HEOM up to sufficiently long real times $t$ such that the integrals determining the carrier mobility [see Eq.~\eqref{Eq:define_mu_dc}] enter saturation as functions of $t$.

The main indicator we use to assess the quality of our HEOM results is the relative accuracy with which the optical sum rule~\cite{PhysRevB.56.4484}
\begin{equation}
\label{Eq:OSR}
    \int_0^{+\infty}d\omega\:\mathrm{Re}\:\mu(\omega)=-\frac{\pi}{2}\langle H_\mathrm{e}+H_\mathrm{e-ph}\rangle
\end{equation}
is satisfied.
Similarly as in our recent HEOM-based study of transport properties of the Holstein model~\cite{JChemPhys.159.094113}, we find that $N$ and $D$ should be chosen sufficiently large so that the relative accuracy
\begin{equation}
    \delta_\mathrm{OSR}=\frac{\left|\int_0^{+\infty}d\omega\:\mathrm{Re}\:\mu(\omega)+\frac{\pi}{2}\langle H_\mathrm{e}+H_\mathrm{e-ph}\rangle\right|}{\frac{\pi}{2}\left|\langle H_\mathrm{e}+H_\mathrm{e-ph}\rangle\right|}
\end{equation} 
becomes essentially independent on $N$ and $D$, and thus mainly determined by the resolution $\Delta\omega$ in the frequency domain.
The spectral resolution $\Delta\omega=\pi/t_\mathrm{max}$ is determined by the maximum time $t_\mathrm{max}$ up to which the hierarchy is propagated (the numerical Fourier transformation is performed on $C_{jj}(t)$ continued to negative times $-t_\mathrm{max}<t<0$).
Quite generally, we find that the convergence of $\langle H_\mathrm{e-ph}\rangle$ with respect to $D$ is slower than the convergence of $\langle H_\mathrm{e}\rangle$ (one example is provided in Sec.~\ref{SSec:bad_example}).
In some situations, we make compromise between minimizing finite-size effects (which requires a sufficiently large $N$) and minimizing errors in $\langle H_\mathrm{e-ph}\rangle$ (which primarily requires a sufficiently large $D$).
In particular, for weaker interactions ($\lambda\lesssim 0.25$) and/or at not too high temperatures ($T/\omega_0\lesssim 5$), when finite-size effects are expected to be pronounced, we sacrifice increasing $D$ to increasing $N$. 
We thus choose $N,D,$ and $t_\mathrm{max}$ sufficiently large so that $\delta_\mathrm{OSR}\lesssim 10^{-3}$.
The present tolerance on $\delta_\mathrm{OSR}$ is an order of magnitude larger than the tolerance we imposed studying the Holstein model~\cite{JChemPhys.159.094113}.
This is not surprising, keeping in mind that $\delta_\mathrm{OSR}$ in the Holstein model is determined only by $\langle H_\mathrm{e}\rangle$, whose convergence with respect to $D$ is controlled better than the convergence of $\langle H_\mathrm{e-ph}\rangle$.

In Sec.~SII of the Supplemental Material~\cite{comment241224}, we establish the equality
\begin{equation}
\label{Eq:cross-contributions-time-reversal}
 \langle j_\mathrm{e}(t)j_\mathrm{e-ph}(0)\rangle-\langle j_\mathrm{e-ph}(t)j_\mathrm{e}(0)\rangle=0   
\end{equation}
as a consequence of the time-reversal symmetry.
In actual computations, Eq.~\eqref{Eq:cross-contributions-time-reversal} is never perfectly satisfied.
We generally find that the maximal magnitude of the LHS of Eq.~\eqref{Eq:cross-contributions-time-reversal} decreases with increasing $D$, while it is not very sensitive to changes in $N$.
The maximal value is generally reached on short time scales $Jt\sim 1$.
We choose $D$ sufficiently large so that the maximal magnitude of the LHS of Eq.~\eqref{Eq:cross-contributions-time-reversal} is of the order of $10^{-2}$ or below.

\section{Numerical results} 
\label{Sec:Numerics}
Here, we explore the viability of the above-introduced HEOM-based approach and study transport properties of the one-dimensional Peierls model.
We focus on the regime of intermediate phonon frequency $\omega_0/J=1$, which has been used to explore practical applicability of various numerically exact methods to the Holstein model~\cite{PhysRevB.100.094307,PhysRevB.102.165155,PhysRevB.105.054311,PhysRevB.106.155129,JChemPhys.159.094113,PhysRevB.107.184315,mitric2024dynamicalquantumtypicalitysimple}.
We also present some results in the antiadiabatic regime of fast phonons $\omega_0/J=3$, in which our hierarchy closing is not entirely effective, see Sec.~\ref{SSec:heom_closing_schemes}.
The results in the adiabatic slow-phonon regime $\omega_0/J\lesssim 0.5$, which are relevant to charge transport in organic semiconductors~\cite{JChemPhys.152.190902,NatMater.19.491,PhysRevLett.96.086601,PhysRevB.83.081202,AdvFunctMater.26.2292}, are presented and discussed in the companion paper~\cite{part2}.
The data that support our conclusions are openly available~\cite{jankovic_2025_14637019}.

As a convenient measure of the electron--phonon interaction strength, we use the dimensionless interaction parameter
\begin{equation}
    \lambda=\frac{2g^2}{\omega_0J}.
\end{equation}
Our choice of $\lambda$ coincides with the definition used in Refs.~\onlinecite{PhysRevB.56.4484,PhysRevLett.105.266605}, and differs from the definition used in Ref.~\onlinecite{AdvFunctMater.26.2292} by a factor of 2.

\subsection{Physical quantities characterizing charge transport}
Although the central quantity of the formalism is $C_{jj}(t)$, see Eq.~\eqref{Eq:def_C_jj}, our time-domain considerations mostly focus on the time-dependent diffusion constant
\begin{equation}
\label{Eq:define_D_t}
    \mathcal{D}(t)=\int_0^t ds\:\mathrm{Re}\:C_{jj}(s).
\end{equation}
In the frequency domain, we examine the dynamical-mobility profile
\begin{equation}
\label{Eq:define_Re_mu_omega}
\begin{split}
    \mathrm{Re}\:\mu(\omega)&=\frac{1-e^{-\beta\omega}}{2\omega}C_{jj}(\omega)\\
    &=\frac{C_{jj}(\omega)-C_{jj}(-\omega)}{2\omega},
\end{split}
\end{equation}
where $C_{jj}(\omega)=\int_{-\infty}^{+\infty}dt\:e^{i\omega t}C_{jj}(t)$.
General symmetries of finite-temperature correlation functions~\cite{Kubo-noneq-stat-mech-book} imply that $C_{jj}(-t)=C_{jj}(t)^*$ and $C_{jj}(-\omega)=e^{-\beta\omega}C_{jj}(\omega)$.
Taking the $\omega\to 0$ limit of Eq.~\eqref{Eq:define_Re_mu_omega} yields the charge mobility
\begin{equation}
\label{Eq:define_mu_dc}
\begin{split}
    \mu_\mathrm{dc}&=\lim_{t\to+\infty}\frac{1}{T}\int_0^{t}ds\:\mathrm{Re}\:C_{jj}(s)\\&=\lim_{t\to+\infty}-2\int_0^{t}ds\:s\:\mathrm{Im}\:C_{jj}(s).
\end{split}
\end{equation}
Equations~\eqref{Eq:define_D_t}--\eqref{Eq:define_mu_dc} imply that the decompositions into band, phonon-assisted, and cross contributions analogous to Eq.~\eqref{Eq:decomposition_C_jj} also hold for $\mathcal{D}(t)$, $\mathrm{Re}\:\mu(\omega)$, and $\mu_\mathrm{dc}$.
The character of the transport is most conveniently discussed in terms of relative magnitudes of different contributions $\mu_\mathrm{dc}^\alpha$ ($\alpha\in\{$e, ph, x$\}$) to carrier mobility.
While $\mu_\mathrm{dc}^\mathrm{e},\mu_\mathrm{dc}^\mathrm{ph}>0$, we find that the cross contribution is negative ($\mu_\mathrm{dc}^\mathrm{x}<0$) in most of the parameter regimes covered.
A convenient measure of the relative importance of the phonon-assisted contribution is
\begin{equation}
\label{Eq:def_S_e-ph_e-ph}
    S_\mathrm{ph}=\frac{\mu_\mathrm{dc}^\mathrm{ph}}{\mu_\mathrm{dc}^\mathrm{e}+\mu_\mathrm{dc}^\mathrm{ph}}.
\end{equation}
As a measure of the relative importance of the cross correlator, we use
\begin{equation}
\label{Eq:def_S_e_e-ph}
 S_\mathrm{x}=\frac{\mu_\mathrm{dc}^\mathrm{x}}{\mu_\mathrm{dc}^\mathrm{e}+\mu_\mathrm{dc}^\mathrm{ph}}.   
\end{equation}

\subsection{Example of a converged calculation}
\label{SSec:good_example}

\begin{figure}
    \centering
    \includegraphics[width=\columnwidth]{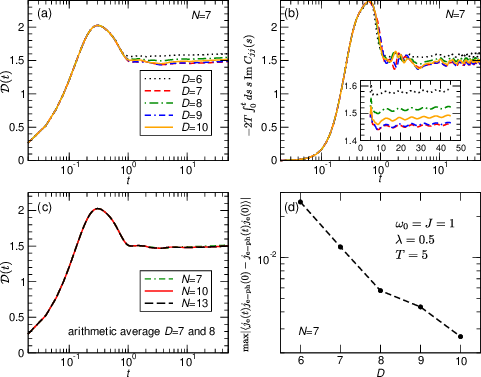}
    \caption{(a) Time-dependent diffusion constant $\mathcal{D}(t)$ computed for $N=7$ and different maximum depths $D$.
    (b) Time evolution of the quantity $-2T\int_0^t ds\:s\:\mathrm{Im}C_{jj}(s)$, which tends to $\mathcal{D}_\infty$ as $t\to+\infty$ [see Eqs.~\eqref{Eq:define_D_t} and~\eqref{Eq:define_mu_dc}].
    The inset shows the same quantity after applying the moving-average procedure described in the text.
    (c) Time-dependent diffusion constant $\mathcal{D}(t)$ computed by averaging the results for $D=7$ and $D=8$ for different chain lengths $N$.
    (d) The maximal modulus of the difference $\langle j_\mathrm{e}(t)j_\mathrm{e-ph}(0)-j_\mathrm{e-ph}(t)j_\mathrm{e}(0)\rangle$ as a function of $D$ for $N=7$.
    The model parameters are $J=\omega_0=1,\lambda=0.5,T=5$.
    The vertical-axis ranges in (a) and (b) are identical.}
    \label{Fig:good_example_031024}
\end{figure}

We find that the herein proposed HEOM-based evaluation of the transport properties of the Peierls model is viable at moderate to high temperatures $T/\omega_0\gtrsim 2$, at which thermally excited phonons are abundant.
It is precisely this temperature range that is relevant to charge transport in high-mobility organic semiconductors, which is ultimately limited by the slow and large-amplitude intermolecular motions~\cite{JChemPhys.152.190902,NatMater.19.491,PhysRevLett.96.086601,PhysRevB.83.081202,AdvFunctMater.26.2292}.
While we discuss the implications of the phonons' slowness (the smallness of the adiabaticity ratio $\omega_0/J$) on charge transport dynamics in the companion paper~\cite{part2}, Figs.~\ref{Fig:good_example_031024}(a)--\ref{Fig:good_example_031024}(d) analyze how finite values of $N$ and $D$ influence the time-dependent diffusion constant for $\omega_0/J=1$, $\lambda=0.5$, and $T/J=5$. 

Fixing $N=7$, we observe that the overall dynamics of $\mathcal{D}$ does not appreciably depend on $D$ as it is varied from 6 to 10, see Fig.~\ref{Fig:good_example_031024}(a).
The same holds for the quantity $-2T\int_0^t ds\:s\:\mathrm{Im}\:C_{jj}(s)$, whose long-time limit should be equal to $\mathcal{D}_\infty$, see Fig.~\ref{Fig:good_example_031024}(b).
We find that using either $\mathrm{Re}\:C_{jj}(t)$ [Fig.~\ref{Fig:good_example_031024}(a)] or $\mathrm{Im}\:C_{jj}(t)$ [Fig.~\ref{Fig:good_example_031024}(b)] yields virtually the same results for $\mathcal{D}_{\infty}$, and thus $\mu_\mathrm{dc}$.
The long-time oscillations observed in Fig.~\ref{Fig:good_example_031024}(b) can be made less pronounced by performing the moving-average procedure~\cite{JChemPhys.159.094113}.
In the inset of Fig.~\ref{Fig:good_example_031024}(b), the result at time $t$ is obtained from the main-panel data by performing the arithmetic average of $N_\mathrm{move}$ main-panel points right before $t$ and $N_\mathrm{move}$ main-panel points right after $t$, where we take $N_\mathrm{move}$ to be 10\% of the total number of data points. 
Both Fig.~\ref{Fig:good_example_031024}(a) and the inset of Fig.~\ref{Fig:good_example_031024}(b) show that the relative variation of $\mathcal{D}_\infty$ upon varying $D$ from 7 to 10 is of the order of 5\%.
For $D\geq 7$, we obtain $\delta_\mathrm{OSR}\sim 10^{-4}$ for the maximum propagation time $Jt_\mathrm{max}=50$.
As discussed in our previous study~\cite{JChemPhys.159.094113}, at sufficiently high temperatures, the convergence with respect to $D$ can be somewhat enhanced by averaging HEOM results for two consecutive depths for which $\delta_\mathrm{OSR}$ is of the same order of magnitude.
We thus conclude that the arithmetic average of the results for $D=7$ and $D=8$ is representative of the result converged with respect to $D$.
In Fig.~\ref{Fig:good_example_031024}(c), we plot $\mathcal{D}(t)$ obtained by averaging HEOM results for $D=7$ and $D=8$ and different values of $N$. 
We find that the finite-size effects are very weakly pronounced, so that HEOM results for $N=7$ are representative of the long-chain limit.
To gain additional confidence in our implementation of the HEOM method, we check how well it respects Eq.~\eqref{Eq:cross-contributions-time-reversal}.
Figure~\ref{Fig:good_example_031024}(d) shows that the maximum of the LHS of Eq.~\eqref{Eq:cross-contributions-time-reversal} over the time interval $[0,t_\mathrm{max}]$ exhibits a slow yet almost exponential decrease with $D$.
For $D=7$ and 8, we see that the maximum is of the order of $10^{-2}$.
While the latter value might seem large, and might suggest that even larger maximum depths are needed to obtain fully converged HEOM results, the results in Figs.~\ref{Fig:good_example_031024}(a)--\ref{Fig:good_example_031024}(d) show that our results are to be regarded as numerically exact for all practical purposes.

In the analysis of the temperature-dependent mobility $\mu_\mathrm{dc}(T)$ in Sec.~\ref{SSec:mu_dc_vs_T}, our final results at temperatures $T/\omega_0\gtrsim 2$ are arithmetic averages of the results obtained using only $\mathrm{Re}\:C_{jj}(t)$ and only $\mathrm{Im}\:C_{jj}(t)$.

\subsection{Challenges at moderate temperatures and interactions. Effectiveness of the hierarchy closing scheme}
\label{SSec:bad_example}

We find it quite challenging to obtain converged results for $\mu_\mathrm{dc}$ at temperatures $T$ such that $T/\omega_0<2$, i.e., when the number of thermally excited phonons is relatively small.

\begin{figure}[htbp!]
    \centering
    \includegraphics[width=\columnwidth]{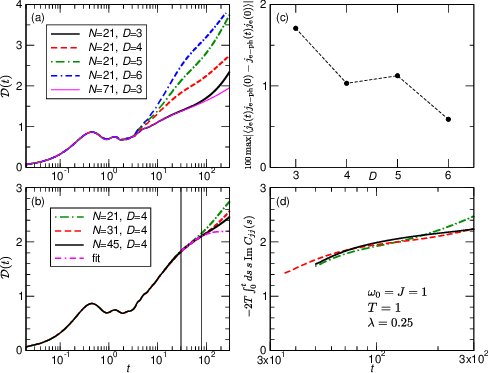}
    \caption{(a) and (b): Time-dependent diffusion constant for (a) $N=21$ and various $D$, (b) $D=4$ and various $N$.
    In (a), we additionally show $\mathcal{D}(t)$ for $N=71,D=3$.
    In (b), the curve labeled "fit" shows the best fit of $\mathcal{D}(t)$ for $N=45$ and $30\leq t\leq 80$ to the exponentially saturating function $f(t)=a_0+a_1\:e^{-t/a_2}$.
    (c) The maximal magnitude of the difference $\langle j_\mathrm{e}(t)j_\mathrm{e-ph}(0)-j_\mathrm{e-ph}(t)j_\mathrm{e}(0)\rangle$ [multiplied by a factor of 100, see Eq.~\eqref{Eq:cross-contributions-time-reversal}] as a function of $D$ for $N=21$.
    (d) The RHS of the second equality in Eq.~\eqref{Eq:define_mu_dc} as a function of $t$ for $D=4$ and different values of $N$.
    The line and color codes in (b) and (d) are identical.
    In all panels, $\omega_0=J=T=1$ and $\lambda=0.25$.}
    \label{Fig:bad_example}
\end{figure}

For moderate to strong interactions $\lambda\gtrsim 0.5$, the numerical instabilities originating from the combination of relatively strong interaction and relatively low temperature prevent us from fully capturing the carrier's diffusive motion and reliably computing the low-frequency dynamical mobility.
The situation is overall similar to that we have encountered in our recent study of the Holstein model, see Sec.~III.F of Ref.~\onlinecite{JChemPhys.159.094113}.
For weak interactions $\lambda\lesssim 0.05$, propagating the hierarchy truncated at $D=1-2$ (with sufficiently large $N$) yields decent results for transport properties, again similarly as in Ref.~\onlinecite{JChemPhys.159.094113}.
The aforementioned challenges, which we have not encountered studying the Holstein model, are the most pronounced for moderate carrier--phonon interactions, e.g., $\lambda=0.25$.
We then find that somewhat larger values of $D$ (typically $3<D<6$) are required to obtain reasonably accurate results for thermodynamic quantities.
This is illustrated in Table~\ref{Tab:e_kin_e_int_J_1_w0_1_lambda_0.25}, which summarizes the results for the carrier's kinetic energy and the carrier--phonon interaction energy for $\omega_0=J=T=1$, $\lambda=0.25$, $N=21$, and different values of $D$.
\begin{table}[htbp!]
    \centering
    \begin{tabular}{c|c|l|c|l}
        $D$ & $-\langle H_\mathrm{e}\rangle$ & $-\langle H_\mathrm{e}\rangle_\mathrm{sig}$ & $-\langle H_\mathrm{e-ph}\rangle$ & $-\langle H_\mathrm{e-ph}\rangle_\mathrm{sig}$ \\\hline
        3 & 1.2163385532 & \textbf{1.216} & 0.74124663880 & \textbf{0.74}\\
        4 & 1.2161795982 & \textbf{1.2161} & 0.74251418518 & \textbf{0.742}\\
        5 & 1.2161720532 & \textbf{1.21617} & 0.74258695586 & \textbf{0.74259}\\
        6 & 1.2161717604 & - & 0.74259024005 & -
    \end{tabular}
    \caption{Carrier's kinetic energy and carrier--phonon interaction energy for $J=\omega_0=T=1$, $\lambda=0.25$, $N=21$ and different values of $D$.
    The timestep on the imaginary axis is set to $\Delta\tau=10^{-2}$.
    Significant figures of the results for $D=3-5$ are reported in bold in separate columns.
    For a quantity $Q_D$ computed using the imaginary-time HEOM truncated at depth $D$, the number of significant figures after the decimal point is the maximum nonnegative integer $n$ satisfying $\left|Q_{D}-Q_{D+1}\right|<5\times 10^{-(n+1)}$.  
    }
    \label{Tab:e_kin_e_int_J_1_w0_1_lambda_0.25}
\end{table}
For each $D$, the significant figures, which are reported in bold in separate columns, are identified by comparing the results at depths $D$ and $D+1$.
We observe that each increase in $D$ by one adds an additional significant figure, and that $\langle H_\mathrm{e-ph}\rangle$ converges more slowly than $\langle H_\mathrm{e}\rangle$.
Table~\ref{Tab:e_kin_e_int_J_1_w0_1_lambda_0.25} suggests that we should set $D\geq 4$ if we want $\delta_\mathrm{OSR}\lesssim 10^{-3}$.
However, Fig.~\ref{Fig:bad_example}(a) shows that $\mathcal{D}(t)$ does not saturate at long real times for $N=21$ and $D=4,5,6$.
Such a behavior, which reflects a very slow long-time decrease of $C_{jj}(t)$ towards zero, could be caused by finite-size effects in the dynamics.
Fixing $D=4$ and increasing $N$ from 21 to 45, we find that the improvement in the long-time behavior of $\mathcal{D}(t)$ is only modest, see Fig.~\ref{Fig:bad_example}(b), and insufficient to reliably estimate $\mu_\mathrm{dc}$.
The same conclusion is reached upon increasing $N$ from 21 to 71 for $D=3$, see Fig.~\ref{Fig:bad_example}(a).
Therefore, the problems we face at moderate temperatures and for moderate interactions originate from the ineffectiveness of our hierarchy closing scheme.

The effectiveness of the closing scheme depends on the model studied, i.e., on the properties of the electron--phonon interaction Hamiltonian.
Namely, the electron--phonon interaction vertex $M(k,q)$ within the Holstein model is independent of both the electron ($k$) and phonon ($q$) momenta, whereas it explicitly depends on both $k$ and $q$ within the present model [Eq.~\eqref{Eq:def_M_k_q}].
The "strength" of the hierarchical links between HEOM auxiliaries is thus independent of the auxiliaries' momenta within the Holstein model, while the links' "strength" within the present model is highly nonuniform due to their pronounced momentum dependence.
In other words, in the Holstein model, the closing-induced hierarchy stabilization is efficiently transferred from the deepest HEOM layer all the way to the HEOM root, thus ensuring the long-time decrease of the current--current correlation function towards zero.
On the other hand, the momentum-dependent hierarchical links in the Peierls model present obstacles to the transfer of the closing-induced hierarchy stabilization towards the HEOM root and its first-layer auxiliaries, thus rendering the decrease of the current--current correlation function slow.
This viewpoint is further corroborated by Fig.~\ref{Fig:bad_example}(a), which shows that increasing $D$ is detrimental to the effectiveness of the closing-induced stabilization.
The larger is the maximum depth, the more abundant are the obstacles due to momentum-dependent hierarchical couplings, and the more inefficient is the stabilization.
This behavior stands in contrast to what we have found in the Holstein model~\cite{JChemPhys.159.094113}, in which increasing $D$ does not appreciably affect the stabilization effectiveness. 

Figure~\ref{Fig:bad_example}(c) shows that the maximal magnitude of the LHS of Eq.~\eqref{Eq:cross-contributions-time-reversal}, which remains $\sim 10^{-2}$ upon varying $D$ from 3 to 6, cannot help us decide on the best value of $D$ (for $N=21$).
Computing the diffusion constant using only $\mathrm{Im}\:C_{jj}(t)$ for $D=4$ and different chain lengths shows that the long-time saturation towards $\mathcal{D}_\infty$ can be inferred from the data for $N=45$.
Almost the same value of $\mathcal{D}_\infty$ can be obtained by fitting the portion of the $\mathcal{D}(t)$ curve for $30\leq Jt\leq 80$ to the exponentially saturating function $f(t)=a_0+a_1\:e^{-t/a_2}$, see the curve labeled "fit" in Fig.~\ref{Fig:bad_example}(b).
The fitting window chosen does not include short-time transients of $\mathcal{D}(t)$, and captures the early approach towards the diffusive transport, during which finite-chain effects are under control.
One might thus regard this fitting procedure to yield $\mathcal{D}(t)$ representative of a currently unaffordable HEOM computation on a longer chain.  
Finally, our result for $\mu_\mathrm{dc}$ is the arithmetic average of the results in Figs.~\ref{Fig:bad_example}(b) and~\ref{Fig:bad_example}(d) at $Jt=300$.
It should be accompanied with the relative uncertainty of the order of $10\%$, which can be estimated from Fig.~\ref{Fig:bad_example}(b) by comparing $\mathcal{D}_\infty$ emerging from the fit and the HEOM data for $N=45$ and $Jt=300$.

\subsection{Influence of the HEOM closing scheme on the results of converged calculations}
\label{SSec:influence_heom_closing_on_converged}

\begin{figure}
    \centering
    \includegraphics[width=\columnwidth]{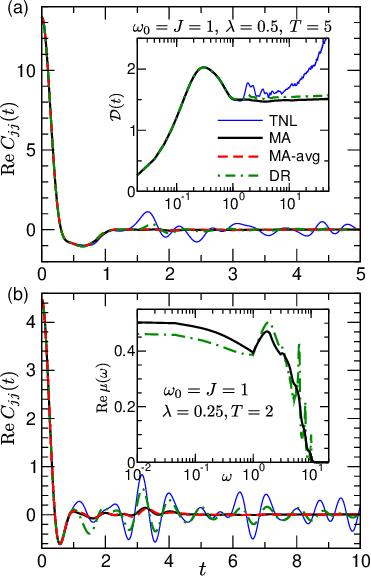}
    \caption{Short-time HEOM dynamics of $\mathrm{Re}\:C_{jj}$ obtained using the TNL [$\Gamma(k,\mathbf{n})\equiv 0$, thin solid line], MA [Eq.~\eqref{Eq:MA_closing}, thick solid line], MA-avg [Eq.~\eqref{Eq:prescription-antiadiabatic}, dashed line], and DR [Eq.~\eqref{Eq:DR_closing_final}, dash-dotted line] closing schemes.
    The insets show (a) the time-dependent diffusion constant and (b) the dynamical-mobility profile.
    For visual clarity, the insets do not show the MA-avg results, which are almost identical to the MA results.
    The model parameters are $J=\omega_0=1$ and (a) $\lambda=0.5,T=5$, (b) $\lambda=0.25,T=2$. 
    Note the logarithmic scale on the horizontal axis of both insets.}
    \label{Fig:complement_pI_fig1}
\end{figure}

Here, we analyze the influence of the HEOM closing schemes introduced in Sec.~\ref{SSec:heom_closing_schemes} on transport dynamics at the high temperature $T/\omega_0=5$ considered in Sec.~\ref{SSec:good_example} [with $\omega_0/J=1$ and $\lambda=0.5$, see Fig.~\ref{Fig:complement_pI_fig1}(a)] and at a lower temperature $T/\omega_0=2$ [with $\omega_0/J=1$ and $\lambda=0.25$, see Fig.~\ref{Fig:complement_pI_fig1}(b)].

Setting $\Gamma(k,\mathbf{n})=0$ in Eq.~\eqref{Eq:re-time-HEOM-final} [the so-called time-nonlocal (TNL) closing~\cite{JChemTheoryComput.8.2808}], the dynamics of $\mathrm{Re}\:C_{jj}$ in Figs.~\ref{Fig:complement_pI_fig1}(a) and~\ref{Fig:complement_pI_fig1}(b) exhibit oscillatory instabilities that prevent us from reliably extracting carrier mobility, see the inset of Fig.~\ref{Fig:complement_pI_fig1}(a).
Notably, these instabilities cannot be removed by further increasing the maximum depth $D$~\cite{JChemPhys.150.184109}.
Both the MA and DR schemes ensure that $\mathrm{Re}\:C_{jj}$ tends to zero at long times, although the MA scheme is more efficient at damping long-time oscillations around zero, especially at the lower temperature considered.
The quantitative differences between $\mathcal{D}_\infty$ or $\mu_\mathrm{dc}$ relying on the MA or DR closing schemes are consistent with the above-established 10\% relative accuracy that should accompany HEOM results, see the appropriate parts of the insets of Figs.~\ref{Fig:complement_pI_fig1}(a) and~\ref{Fig:complement_pI_fig1}(b).

At sufficiently high temperatures and for sufficiently strong interactions, as in Fig.~\ref{Fig:complement_pI_fig1}(a), the diffusive transport typically sets in before the appearance of oscillatory instabilities in the TNL-closing solution.
Then, stopping HEOM propagation before the TNL-closing instabilities have developed [at $Jt\approx 1$ in Fig.~\ref{Fig:complement_pI_fig1}(a)], all four closing possibilities considered yield virtually the same value for the carrier mobility, see the inset of Fig.~\ref{Fig:complement_pI_fig1}(a).
Still, the full dynamical-mobility profile with a decent spectral resolution necessitates dynamics up to longer times, see Sec.~\ref{SSec:technical-details}.
The explicit long-time propagation can be avoided by resorting to, e.g., zero padding~\cite{PhysRevB.106.155129}, which is justified in the situation analyzed in Fig.~\ref{Fig:complement_pI_fig1}(a).
However, the low-frequency features in $\mathrm{Re}\:\mu(\omega)$ stemming from the zero-padded signal may be unreliable.
Such features can be of interest in realistic situations, see the companion paper~\cite{part2}, and we prefer explicit propagation of the HEOM with an appropriate closing scheme to techniques such as zero padding.

At lower temperatures, there are some qualitative differences between the dynamical-mobility profiles relying on the MA and DR closing schemes, see the inset of Fig.~\ref{Fig:complement_pI_fig1}(b).
Namely, the dynamical mobility relying on the DR closing displays pronounced high-frequency features stemming from the short-time oscillatory features that closely replicate TNL-closing instabilities, see Fig.~\ref{Fig:complement_pI_fig1}(b).
Such features remain appreciable even for weaker interactions, at which the Boltzmann transport theory is plausible.
The Boltzmann theory yields Drude-like dynamical-mobility profiles, which are smooth at high frequencies.
For weak interactions, the MA closing yields such smooth dynamical-mobility profiles, see Fig.~\ref{Fig:details_rearranged_071024}(a2), and carrier mobilities that agree very well with the Boltzmann theory, see Fig.~\ref{Fig:heom-vs-boltzmann_071024}.
Therefore, we give preference to the MA over the DR closing scheme at lower temperatures.

Overall, we conclude that the MA closing scheme stabilizes HEOM dynamics in a manner that does not introduce spurious high-frequency features and does not compromise low-frequency features of the dynamical mobility, and, in particular, the magnitude of the carrier mobility.  

The $k$-independent version [Eq.~\eqref{Eq:prescription-antiadiabatic}] of the MA closing term yields virtually the same dynamics as the MA closing term, compare the curves "MA" and "MA-avg" in Figs.~\ref{Fig:complement_pI_fig1}(a) and~\ref{Fig:complement_pI_fig1}(b).
This strongly suggests that our manner of enhancing the effectiveness of the MA closing for $\omega_0/J\geq 2$ does not introduce additional artifacts into HEOM dynamics.

\subsection{Temperature-dependent charge mobility}
\label{SSec:mu_dc_vs_T}

\begin{figure*}[htbp!]
    \centering
    \includegraphics[width=\textwidth]{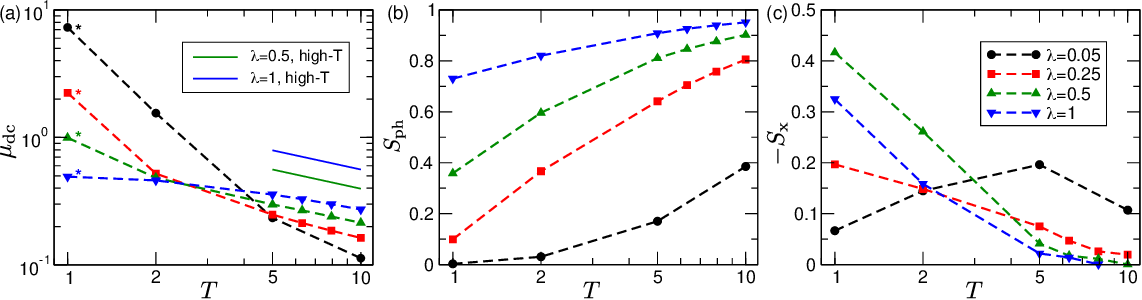}
    \caption{Temperature dependence of (a) the carrier mobility $\mu_\mathrm{dc}$, (b) the share $S_\mathrm{ph}$ of the phonon-assisted contribution to $\mu_\mathrm{dc}$ [Eq.~\eqref{Eq:def_S_e-ph_e-ph}], and (c) the share $-S_\mathrm{x}$ of the cross contribution to $\mu_\mathrm{dc}$ [Eq.~\eqref{Eq:def_S_e_e-ph}] for different carrier--phonon interaction strengths.
    In all panels, $\omega_0=J=1$.
    The results at $T=1$, which are accompanied by asterisks, may not be fully converged.
    Panel (a) additionally displays the results of Eq.~\eqref{Eq:mu_dc_high-T} when the transport is predominantly phonon-assisted ($\lambda=0.5$ and 1, $5\leq T\leq 10$).
    }
    \label{Fig:mu_vs_T_omega_1_overall_021024}
\end{figure*}

Figure~\ref{Fig:mu_vs_T_omega_1_overall_021024}(a) summarizes our results for the temperature-dependent carrier mobility for $\omega_0/J=1$ and different interaction strengths.
Section~SIII of the Supplemental Material~\cite{comment241224} summarizes the parameter regimes in which we performed HEOM computations, along with the corresponding numerical parameters ($N$, $D$, and the maximum propagation time $t_\mathrm{max}$).
Sections~\ref{SSec:good_example} and~\ref{SSec:bad_example} show that the relative uncertainties accompanying $\mu_\mathrm{dc}(T)$ in Fig.~\ref{Fig:mu_vs_T_omega_1_overall_021024}(a) generally decrease with $T$, and are of the order of (or somewhat below) $10\%$ throughout the temperature range examined. 

Fixing $\lambda$, we find that $\mu_\mathrm{dc}$ decreases with temperature within the temperature range examined. 
This decrease becomes milder at higher temperatures and/or for stronger interactions.
At $T/J=1$, $\mu_\mathrm{dc}$ decreases with increasing interaction, while the opposite trend is observed at temperatures $T/J=5$ and 10.
We connect these findings with the character of the transport in Figs.~\ref{Fig:mu_vs_T_omega_1_overall_021024}(b) and~\ref{Fig:mu_vs_T_omega_1_overall_021024}(c), which respectively present the temperature dependence of the phonon-assisted [Eq.~\eqref{Eq:def_S_e-ph_e-ph}] and cross [Eq.~\eqref{Eq:def_S_e_e-ph}] shares of the mobility.
We conclude that the opposite trends in $\mu_\mathrm{dc}$ with increasing interaction reflect the crossover from the transport dominated by the purely electronic contribution at lower temperatures towards the phonon-assisted transport at higher temperatures.
Figure~\ref{Fig:mu_vs_T_omega_1_overall_021024}(a) suggests that the crossover takes place at temperatures around $2J$, at which $\mu_\mathrm{dc}$ is almost independent of the interaction as long as it is sufficiently strong ($\lambda\gtrsim 0.25$). 
In contrast to the phonon-assisted contribution, which gains importance as the temperature and/or the interaction are increased, see Fig.~\ref{Fig:mu_vs_T_omega_1_overall_021024}(b), the cross contribution is the most appreciable for moderate interactions and/or at lower temperatures, see Fig.~\ref{Fig:mu_vs_T_omega_1_overall_021024}(c).

\begin{figure}[htbp!]
    \centering
    \includegraphics[width=.9\columnwidth]{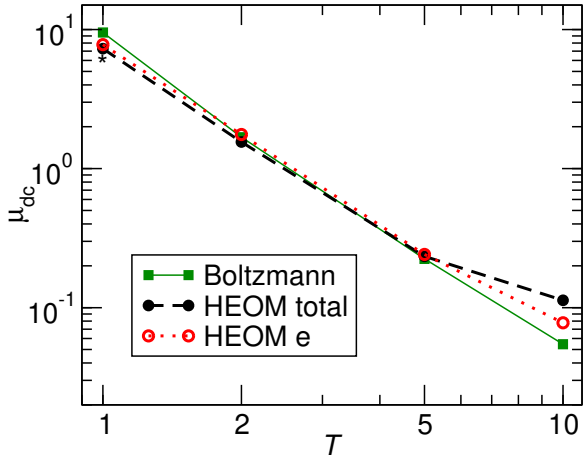}
    \caption{Temperature-dependent charge mobility computed using the HEOM method (label "HEOM total") and the Boltzmann equation (label "Boltzmann", see Appendix~\ref{App:Boltzmann}).
    We also show the HEOM results for the band contribution to $\mu_\mathrm{dc}$ (label "HEOM e").
    The model parameters are $\omega_0=J=1$ and $\lambda=0.05$. The data labeled "Boltzmann" are the courtesy of N. Vukmirovi\'c.}
    \label{Fig:heom-vs-boltzmann_071024}
\end{figure}

The above-described trends in $\mu_\mathrm{dc}$ upon varying $T$ and $\lambda$ in the regime of predominantly phonon-assisted transport can be reproduced by the early theories developed in Refs.~\cite{PhysRev.150.529,JChemPhys.70.3775}.
Assuming that the temperature is the largest energy scale in the problem, factorizing carrier--phonon correlators as products of purely carrier and free-phonon correlators, and computing the former in the independent-particle (bubble) approximation using the local (momentum-independent) carrier propagator (as in, e.g., Sec.~III.C of Ref.~\onlinecite{PhysRevB.109.214312}), one arrives at (see also Sec.~3.1 of Ref.~\onlinecite{AdvFunctMater.26.2292})  
\begin{equation}
\label{Eq:mu_dc_high-T}
    \mu_\mathrm{dc}^{\mathrm{high}-T}=\sqrt{\frac{\pi}{4\lambda}}\left(\frac{J}{T}\right)^{3/2}\left(1+2\lambda\frac{T}{J}\right)\approx\sqrt{\pi\lambda\frac{J}{T}}.
\end{equation}
Figure~\ref{Fig:mu_vs_T_omega_1_overall_021024}(a) reveals that Eq.~\eqref{Eq:mu_dc_high-T}, which predicts $\mu_\mathrm{dc}^{\mathrm{high}-T}\propto T^{-0.5}$, reasonably reproduces the exponent of the power-law decrease of the numerically exact $\mu_\mathrm{dc}$ with $T$ for $\lambda\gtrsim 0.5$ and at $T/J\gtrsim 5$.
The fits of the HEOM results for $\mu_\mathrm{dc}(T)$ to a power-law function are performed in Sec.~SIV of the Supplemental Material~\cite{comment241224}.  
Figure~\ref{Fig:mu_vs_T_omega_1_overall_021024}(a) also suggests that the dependence of the HEOM mobility on $\lambda$ for fixed $T$ is weaker than predicted by Eq.~\eqref{Eq:mu_dc_high-T}.
Importantly, Eq.~\eqref{Eq:mu_dc_high-T} severely overestimates the numerically exact results, which can be traced back to the bubble approximation inherent to Eq.~\eqref{Eq:mu_dc_high-T}~\cite{AdvFunctMater.26.2292}.

For $\lambda=0.05$, we expect that the mobility within the Boltzmann transport theory~\cite{Mahanbook,ALLEN2006165,Jacoboni}, which considers only the purely electronic contribution to $\mu_\mathrm{dc}$, should closely follow HEOM results at least at lower temperatures featuring small phonon-assisted and cross contributions.
This expectation in confirmed in Fig.~\ref{Fig:heom-vs-boltzmann_071024}, which compares the predictions $\mu_\mathrm{dc}^\mathrm{Bltz}$ of the Boltzmann theory with the (total) HEOM mobility $\mu_\mathrm{dc}$ (label "HEOM total") and its purely electronic contribution $\mu_\mathrm{dc}^\mathrm{e}$ (label "HEOM e").
Figure~\ref{Fig:heom-vs-boltzmann_071024} shows that the Boltzmann theory accurately reproduces HEOM results up to temperatures $T/J\sim 5$.
Interestingly, although the phonon-assisted and cross contributions to mobility are both sizable at $T/J=5$, these approximately cancel one another, see Figs.~\ref{Fig:mu_vs_T_omega_1_overall_021024}(b) and~\ref{Fig:mu_vs_T_omega_1_overall_021024}(c), so that the $\mu_\mathrm{dc}^\mathrm{Bltz}$ is almost identical to the numerically exact mobility.
At $T/J=10$, the Boltzmann theory underestimates already the purely electronic contribution to the mobility.
Still, the deviation of its prediction from the total HEOM result is mainly due to the considerable phonon-assisted contribution.
We note that the Boltzmann results presented in Fig.~\ref{Fig:heom-vs-boltzmann_071024} go beyond the usually employed approximations, such as the momentum relaxation time approximation~\cite{RepProgPhys.83.036501,PhysRevResearch.2.013001}.
We discuss this aspect in greater detail in Appendix~\ref{App:Boltzmann}.

For $\lambda=1$, the temperature dependence of $\mu_\mathrm{dc}$ is very weak at the lower end of the temperature range examined, while it can be reasonably approximated by $\mu_\mathrm{dc}\propto T^{-0.5}$ at the higher end of that range, see Fig.~\ref{Fig:mu_vs_T_omega_1_overall_021024}(a) and Sec.~SIV of the Supplementary Material~\cite{comment241224}.
An overall similar behavior of $\mu_\mathrm{dc}(T)$ was observed in Ref.~\onlinecite{PhysRevB.69.075212} for sufficiently strong interactions (see the temperature-dependent dc conductivity labeled "$\sigma_c^\mathrm{DC}$" in Fig.~1 of Ref.~\onlinecite{PhysRevB.69.075212}).
The almost temperature-independent mobility we observe upon decreasing the temperature to $T/J\sim 1$ most probably corresponds to the well studied thermally activated transport, in which the mobility weakly increases with temperature~\cite{JChemPhys.70.3775}.
This type of transport characterizes the transition from the low-temperature band transport, for which $\mu_\mathrm{dc}\propto T^{-1}$~\cite{PhysRevB.99.104304}, to the high-temperature phonon-assisted transport~\cite{PhysRevB.69.075212}, and is also observed in the Holstein model~\cite{PhysRevB.99.104304}.

\begin{figure*}[htbp!]
    \centering
    \includegraphics[width=\textwidth]{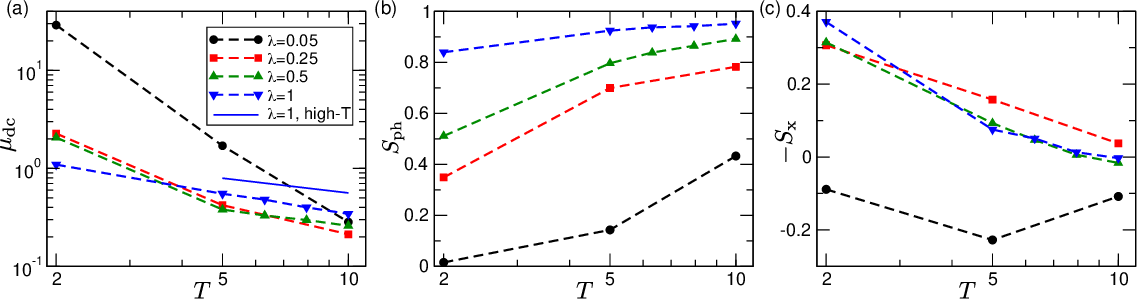}
    \caption{Temperature dependence of (a) the carrier mobility $\mu_\mathrm{dc}$, (b) the share $S_\mathrm{ph}$ of the phonon-assisted contribution to $\mu_\mathrm{dc}$ [Eq.~\eqref{Eq:def_S_e-ph_e-ph}], and (c) the share $-S_\mathrm{x}$ of the cross contribution to $\mu_\mathrm{dc}$ [Eq.~\eqref{Eq:def_S_e_e-ph}] for different carrier--phonon interaction strengths.
    In all panels, $J=1$ and $\omega_0=3$.
    The results may not be entirely reliable because our HEOM closing scheme is only partially effective for $\omega_0/J\geq 2$.
    Panel (a) additionally displays the results of Eq.~\eqref{Eq:mu_dc_high-T} for $\lambda=1$ and at $5\leq T\leq 10$.
    }
    \label{Fig:mu_vs_T_omega_3_overall_241024}
\end{figure*}

Finally, we discuss the temperature-dependent carrier mobility in the fast-phonon regime $\omega_0/J=3$ in Figs.~\ref{Fig:mu_vs_T_omega_3_overall_241024}(a)--\ref{Fig:mu_vs_T_omega_3_overall_241024}(c).
Overall, the trends displayed by the mobility and its contributions upon variations in $T$ and $\lambda$ are similar to those in Fig.~\ref{Fig:mu_vs_T_omega_1_overall_021024}(a).
Interestingly, in contrast to our findings in Fig.~\ref{Fig:mu_vs_T_omega_1_overall_021024}(c), in Fig.~\ref{Fig:mu_vs_T_omega_3_overall_241024}(c) we find that the cross contribution to $\mu_\mathrm{dc}$ is positive for weak interactions.
Comparing Figs.~\ref{Fig:mu_vs_T_omega_3_overall_241024}(b) and~\ref{Fig:mu_vs_T_omega_1_overall_021024}(b), we find that the shares of the phonon-assisted contribution at $T=\frac{2}{3}\omega_0$ and $T=\frac{5}{3}\omega_0$ in Fig.~\ref{Fig:mu_vs_T_omega_3_overall_241024}(b) are greater than the shares at $T=\omega_0$ and $T=2\omega_0$ in Fig.~\ref{Fig:mu_vs_T_omega_1_overall_021024}(b), respectively.
This observation suggests that faster phonons promote a faster transition to the predominantly phonon-assisted transport with increasing interaction at a fixed temperature.
A similar trend is observed upon increasing temperature at a fixed interaction.
Figure~\ref{Fig:mu_vs_T_omega_3_overall_241024}(a) suggests that the decrease of the mobility for $T\gtrsim\omega_0$ and sufficiently strong interaction can be reasonably approximated by the power-law $\mu_\mathrm{dc}\propto T^{-\alpha}$ with $\alpha\sim 0.5$, as predicted by Eq.~\eqref{Eq:mu_dc_high-T}.
The corresponding fits are provided in Sec.~SIV of the Supplemental Material~\cite{comment241224}.

\subsection{Insights from time and frequency domain}

\begin{figure}[htbp!]
    \centering
    \includegraphics[width=\columnwidth]{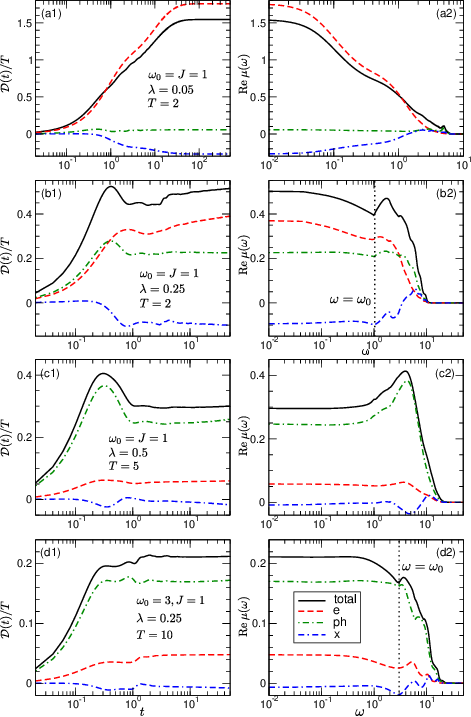}
    \caption{Purely electronic (dashed lines, label "e"), phonon-assisted (dash-dotted lines, label "ph"), and cross (double dash-dotted lines, label "x") contributions to: (a1)--(d1) the time-dependent diffusion constant (normalized by the temperature); (a2)--(d2) the dynamical-mobility profile.
    Solid lines (label "total") show the sum of the three contributions.
    The values of model parameters are cited inside panels (a1)--(d1).}
    \label{Fig:details_rearranged_071024}
\end{figure}

In Fig.~\ref{Fig:details_rearranged_071024}, we analyze how the relative importance of different contributions to transport affects the signatures of the carrier--phonon interaction in the time [$\mathcal{D}(t)$, Figs.~\ref{Fig:details_rearranged_071024}(a1)--\ref{Fig:details_rearranged_071024}(d1)] and frequency [$\mathrm{Re}\:\mu(\omega)$, Figs.~\ref{Fig:details_rearranged_071024}(a2)--\ref{Fig:details_rearranged_071024}(d2)] domains.
As the infinite-time limit of $\mathcal{D}(t)/T$ and the zero-frequency limit of $\mathrm{Re}\:\mu(\omega)$ should coincide, Figs.~\ref{Fig:details_rearranged_071024}(a1)--\ref{Fig:details_rearranged_071024}(d1) show $\mathcal{D}(t)$ normalized by the temperature. 
We start with a regime in which the purely electronic contribution dominates transport properties, see Figs.~\ref{Fig:details_rearranged_071024}(a1) and~\ref{Fig:details_rearranged_071024}(a2), and proceed by increasing the relative importance of the phonon-assisted contribution, see Figs.~\ref{Fig:details_rearranged_071024}(b1)--\ref{Fig:details_rearranged_071024}(c2).
Figures~\ref{Fig:details_rearranged_071024}(d1) and~\ref{Fig:details_rearranged_071024}(d2) present some reliable results for fast phonons.

When the band contribution dominates the transport, the diffusion constant steadily increases with time, see Fig.~\ref{Fig:details_rearranged_071024}(a1), and the dynamical-mobility profile is overall Drude-like, see Fig.~\ref{Fig:details_rearranged_071024}(a2).
The Drude-like shape of the purely electronic contribution to $\mathrm{Re}\:\mu(\omega)$ is not affected by the nontrivial cross contribution, which simply provides a nonuniform (in frequency) shift of the purely electronic contribution in the low-frequency region.
Even though the phonon-assisted contribution is negligible, it leaves its footprint in the high-frequency region in form of a low-intensity peak centered around $\omega/J=5$.
Keeping in mind the definition of the phonon-assisted current in Eqs.~\eqref{Eq:def_j_e-ph}--\eqref{Eq:def_M_J_k_q}, this peak most probably reflects a highly off-resonant process in which a single phonon 
belonging to the totally antisymmetric mode ($q=\pi$) mediates transitions between the bottom ($k=0,\varepsilon_{k=0}=-2J$) and the top ($k=\pi,\varepsilon_{k=\pi}=2J$) of the electronic band.
These values of $k$ and $q$ maximize the matrix element $M_J(k,q)$ in Eq.~\eqref{Eq:def_M_J_k_q}, which renders the corresponding peak visible.

When the band and phonon-assisted contributions are comparable, the diffusion constant displays nonmonotonic behavior that is qualitatively similar to that we have studied in the Holstein model~\cite{JChemPhys.159.094113,PhysRevB.109.214312}, see Fig.~\ref{Fig:details_rearranged_071024}(b1).
After the initial ballistic-like increase, the diffusion constant exhibits a decrease after $Jt\sim 0.4$, which is mainly due to the phonon-assisted contribution.
While the purely electronic contribution to $\mathcal{D}$ exhibits a similar behavior on these timescales, it reaches its maximum somewhat later than the phonon-assisted contribution.
Finally, on longer timescales, the phonon-assisted contribution to $\mathcal{D}$ decreases and saturates, whereas the purely electronic contribution to $\mathcal{D}$ increases.
The dynamical-mobility profile in Fig.~\ref{Fig:details_rearranged_071024}(b2) displays a local minimum around $\omega=\omega_0$ and a broad finite-frequency peak.
The weakly pronounced, yet observable, long-time growth of $\mathcal{D}(t)$ in Fig.~\ref{Fig:details_rearranged_071024}(b1) can be traced back to the temperature $T/\omega_0=2$ lying on the borderline between the regions in which our computational framework is (un)feasible.

The prevalence of the phonon-assisted contribution is characterized by the diffusion constant that exhibits no long-time increase, but approaches its long-time limit while decreasing, see Fig.~\ref{Fig:details_rearranged_071024}(c1).
The dynamical-mobility profile is then dominated by a finite-frequency peak, see Fig.~\ref{Fig:details_rearranged_071024}(c2), and exhibits a local minimum at $\omega=0$.
This stands in contrast to Fig.~\ref{Fig:details_rearranged_071024}(b2), in which $\omega=0$ is a local maximum in $\mathrm{Re}\:\mu(\omega)$.
Overall, the results in Figs.~\ref{Fig:details_rearranged_071024}(c1) and~\ref{Fig:details_rearranged_071024}(c2) bear qualitative resemblance to typical predictions of the transient localization scenario (TLS)~\cite{PhysRevB.83.081202,AdvFunctMater.26.2292}.
However, we emphasize that here $\omega_0/J=1$, while the TLS is physically plausible in the limit of slow phonons.
In the companion paper~\cite{part2}, we present our HEOM results for small adiabaticity ratios and assess the appropriateness of the TLS.

When the phonon-assisted contribution dominates the transport for $\omega_0/J=3$, the behavior of $\mathcal{D}(t)$ and $\mathrm{Re}\:\mu(\omega)$ is overall Drude-like, see Figs.~\ref{Fig:details_rearranged_071024}(d1) and~\ref{Fig:details_rearranged_071024}(d2).
Similarly as in Fig.~\ref{Fig:details_rearranged_071024}(b1), we find that the phonon-assisted contribution to $\mathcal{D}$ exhibits no long-time increase, in contrast to the purely electronic contribution to $\mathcal{D}$, which increases in a step-like fashion.
Such a behavior of the purely electronic contribution to $\mathcal{D}$ qualitatively resembles our findings within the Holstein model~\cite{JChemPhys.159.094113}.
Although the position of the steps is not seemingly correlated with the integer multiples of the phonon period, the dynamical-mobility profile in Fig.~\ref{Fig:details_rearranged_071024}(d2) exhibits a dip around $\omega=\omega_0$, which is mainly due to the purely electronic contribution, cf. Fig.~\ref{Fig:details_rearranged_071024}(b2).

\section{Summary and outlook}
\label{Sec:Summary}
In this study, we have overcome the long-standing challenge of correctly treating the phonon-assisted current in HEOM-based computations of transport properties of models with nonlocal carrier--phonon interaction.
Admittedly, the general ideas needed to address the challenge have been developed in different setups~\cite{JChemPhys.145.224105,PhysRevB.95.064308,PhysRevB.97.235429,ChemPhys.515.129}, and ultimately systematized in the DEOM formalism~\cite{MolPhys.116.780,JChemPhys.157.170901}.
However, our approach combines them in a novel manner, shedding new light on the very nature of the HEOM formalism and its dynamical variables in one particular case.
It is 
a model in which a charge carrier moving on a lattice interacts with an environment composed of a finite number of undamped harmonic oscillators, which has received much attention in different contexts~\cite{PhysRevLett.96.086601,PhysRevB.83.081202,AdvFunctMater.26.2292,JChemPhys.150.184109,JChemPhys.153.204109,JChemPhys.160.111102,arXiv.2406.19851}.

We explicitly express HEOM auxiliaries in terms of single oscillator's creation and annihilation operators [Eqs.~\eqref{Eq:ADM_multiphonon_config} and~\eqref{Eq:F_n_normal_ordered}].
The auxiliaries are found to describe many phonon-assisted transitions between free-carrier states that are mediated by genuine many-phonon correlations, from which the redundant information already present in lower-order auxiliaries is eliminated [Eq.~\eqref{Eq:F_n_F_n}].
We then rigorously prove the generalized Wick's theorem [Eqs.~\eqref{Eq:GWT_left_DEOM} and~\eqref{Eq:GWT_right_DEOM}], which is the essential ingredient of the computational framework (Sec.~\ref{Sec:HEOM_Peierls}) that handles finite-temperature correlation functions of mixed carrier--phonon operators.

This framework is then used to obtain numerically exact transport properties of the one-dimensional Peierls model.
We find that our HEOM-based computations deliver reliable results for the carrier mobility only when phonons are abundantly thermally excited.
At sufficiently high temperatures, and for sufficiently strong interactions, thermal fluctuations of the carrier transfer integral become so pronounced that they provide the main driving force for the long-distance carrier transport.
The phonon-assisted nature of transport can be inferred from the prevalence of the phonon-assisted over the band contribution to carrier mobility, so that the mobility increases with interaction at a fixed temperature.
Another indicator confirming that the transport is phonon-assisted is the temperature dependence of the mobility, which, for fixed interaction, follows the power-law behavior $\mu_\mathrm{dc}(T)\propto T^{-\alpha}$ with $\alpha\approx 0.5$.
Our results suggest that the minimum interaction and temperature above which the transport can be considered as phonon-assisted decrease as phonon dynamics becomes faster with respect to carrier dynamics.
The pronounced displaced Drude peak in the carrier's optical response reflects the predominance of the phonon-assisted transport channel when the timescales of free-carrier and free-phonon dynamics are comparable.

While here we have focused on intermediate to fast phonons, our computational framework lends itself to providing the long-awaited quantum dynamical insights into the fundamentals of carrier transport in the field of slow, large-amplitude intermolecular phonons.
The corresponding physical situation, which is relevant to transport in mechanically soft semiconductors, is analyzed in the companion paper~\cite{part2}.
Our methodological developments could motivate further studies concerning the fundamentals of quantum dissipation.
In the language of the DEOM formalism, we have obtained explicit expressions for single dissipatons and many-dissipaton configurations in a model in which the environment is not a real and proper bath, i.e., in which the dissipation is not apparent.
The ideas proposed here could be useful in more explicitly connecting the quasiparticle picture of dissipation embodied in the DEOM formalism with the microscopic bath Hamiltonian, and clarifying the pathway from the reversible system-plus-bath dynamics towards the irreversible system dynamics. 

\begin{acknowledgments}
This research was supported by the Science Fund of the Republic of Serbia, Grant No. 5468, Polaron Mobility in Model Systems and Real Materials--PolMoReMa.
The author acknowledges funding provided by the Institute of Physics Belgrade through a grant from the Ministry of Science, Technological Development, and Innovation of the Republic of Serbia.
Numerical computations were performed on the PARADOX-IV supercomputing facility at the Scientific Computing Laboratory, National Center of Excellence for the Study of Complex Systems, Institute of Physics Belgrade.
The author thanks Nenad Vukmirovi\'c for many useful and stimulating discussions.
\end{acknowledgments}

\section*{Data Availability Statement}
The data that support the findings of this article are openly available~\cite{jankovic_2025_14637019}.

\newpage

\begin{widetext}
\appendix
\section{HEOM formal definitions}
\label{App:HEOM_formal}
According to the Feynman--Vernon influence functional theory~\cite{AnnPhys.24.118}, the only phononic quantity influencing the reduced carrier dynamics in Eq.~\eqref{Eq:define_rdm_t} is $\mathcal{C}_{q_2q_1}(t)=\left\langle B_{q_2}^{(I)}(t)B_{q_1}^{(I)}(0)\right\rangle_\mathrm{ph}$ ($t>0$), which is proportional to the greater free-phonon Green's function.
The lesser counterpart of this quantity is $\left\langle B_{q_1}^{(I)}(0)B_{q_2}^{(I)}(t)\right\rangle_\mathrm{ph}=\mathcal{C}_{\overline{q_2}\:\overline{q_1}}(t)^*$.
Time-dependent operators in the interaction picture are defined as $O^{(I)}(t)=e^{i(H_\mathrm{e}+H_\mathrm{ph})t}Oe^{-i(H_\mathrm{e}+H_\mathrm{ph})t}$.
The construction of the hierarchically coupled equations [Eq.~\eqref{Eq:HEOM-original}] relies on the exponential decompositions ($t>0$)
\begin{equation}
\label{Eq:free-phonon-autocorrelation}
    \mathcal{C}_{q_2q_1}(t)=\sum_m\eta_{q_2q_1m}e^{-\mu_mt},\quad \mathcal{C}_{\overline{q_2}\:\overline{q_1}}(t)^*=\sum_m\eta_{\overline{q_2}\:\overline{q_1}\:\overline{m}}^*\:e^{-\mu_m t},
\end{equation}
where we introduce $\overline{m}$ by $\mu_{\overline{m}}=\mu_m^*$.
For the model specified in Eqs.~\eqref{Eq:H_tot}--\eqref{Eq:def_B_q}, the sums in Eq.~\eqref{Eq:free-phonon-autocorrelation} contain two terms ($m=0,1$), and the quantities $\eta_{q_2q_1m}$ and $\mu_m$ read
\begin{eqnarray}
    \eta_{q_2q_10}=\delta_{q_1\overline{q_2}}\:c_0,\: c_0=\frac{\left(\frac{g}{\sqrt{N}}\right)^2}{1-e^{-\beta\omega_0}},\:\mu_0=i\omega_0,\label{Eq:def_c_0}\\
    \eta_{q_2q_11}=\delta_{q_1\overline{q_2}}\:c_1,\: c_1=\frac{\left(\frac{g}{\sqrt{N}}\right)^2}{e^{\beta\omega_0}-1},\quad\mu_1=-i\omega_0.\label{Eq:def_c_1}
\end{eqnarray}
In our previous studies~\cite{PhysRevB.105.054311,JChemPhys.159.094113,PhysRevB.109.214312}, we incorporated the momentum conservation [embodied in the factor $\delta_{q_1\overline{q_2}}$ entering Eqs.~\eqref{Eq:def_c_0} and~\eqref{Eq:def_c_1}] into the formalism from the outset, and omitted the complex conjugation in Eq.~\eqref{Eq:free-phonon-autocorrelation}, which is justified in models with undamped phonons.
Here, keeping the formalism as general as possible facilitates our formal developments, and reveals their connections to the DEOM formalism~\cite{MolPhys.116.780,JChemPhys.157.170901}.

Equation~\eqref{Eq:HEOM-original} is most conveniently derived by assuming that the interacting electron--phonon system starts from the factorized initial condition $\rho_\mathrm{tot}(0)=\rho(0)\rho_\mathrm{ph}^\mathrm{eq}$ in Eq.~\eqref{Eq:define_rdm_t}.
Then, the electronic RDM at instant $t$ and in the interaction picture is~\cite{JChemPhys.130.234111,JChemPhys.153.244122,JChemPhys.159.094113,PhysRevB.109.214312} 
\begin{equation}
    \rho^{(I)}(t)=\mathcal{T}e^{-\Phi(t)}\rho(0),
\end{equation}
where $\mathcal{T}$ denotes the chronological time-ordering sign (latest superoperator to the left), while
\begin{equation}
\label{Eq:def_Phi_t}
    \Phi(t)=i\sum_{qm}\int_0^t ds\:V_{q}^{(I)}(s)^\times
    \varphi_{qm}(s).
\end{equation}
The superoperators $V^\times$ and $V^\circ$ act on an arbitrary operator $O$ as $V^\times O=[V,O]$ (commutator) and $V^\circ O=\{V,O\}$ (anticommutator). 
The superoperator
\begin{equation}
\label{Eq:def_varphi_qm_t}
\begin{split}
    &\varphi_{qm}(s)=-i\sum_{q'}\int_0^s ds'\:e^{-\mu_m(s-s')}\times\\
    &\left[\frac{\eta_{qq'm}+\eta_{\overline{q}\:\overline{q'}\:\overline{m}}^*}{2}V_{q'}^{(I)}(s')^\times+\frac{\eta_{qq'm}-\eta_{\overline{q}\:\overline{q'}\:\overline{m}}^*}{2}V_{q'}^{(I)}(s')^\circ\right]
\end{split}
\end{equation}
defines the interaction-picture auxiliaries $\rho_\mathbf{n}^{(n,I)}(t)$ as
\begin{equation}
\label{Eq:define-adm-fvift}
   \rho_\mathbf{n}^{(n,I)}(t)=\mathcal{T}\prod_{qm}\varphi_{qm}(t)^{n_{qm}}e^{-\Phi(t)}\rho(0).
\end{equation}
Equation~\eqref{Eq:define-adm-fvift} uses the second-quantization-like definition of vector $\mathbf{n}$, see Eq.~\eqref{Eq:vector_n_number_representation}.
Together with Eq.~\eqref{Eq:def_varphi_qm_t}, it reveals that
\begin{equation}
    \mu_\mathbf{n}=\sum_{qm}n_{qm}\mu_m=i\omega_0\sum_q\left(n_{q0}-n_{q1}\right).
\end{equation}
If we adopt the first-quantization-like definition in Eq.~\eqref{Eq:vector_n_momentum_representation}, we rewrite Eq.~\eqref{Eq:define-adm-fvift} as
\begin{equation}
\label{Eq:define-adm-fvift-momentum-representation}
    \rho_\mathbf{n}^{(n,I)}(t)=\mathcal{T}\prod_{a=1}^n\varphi_{q_am_a}(t)e^{-\Phi(t)}\rho(0).
\end{equation}

We emphasize that the hierarchical structure of Eq.~\eqref{Eq:HEOM-original} is independent on the particular form of $\rho_\mathrm{tot}(0)$, which only determines the initial conditions $\rho_\mathbf{n}^{(n)}(0)$ for HEOM auxiliaries~\cite{JChemPhys.141.044114}.

\section{Connecting HEOM auxiliaries and many-phonon-assisted processes: Derivation of Eqs.~\eqref{Eq:F_n_normal_ordered} and~\eqref{Eq:F_n_F_n}}
\label{App:Schwinger}

The derivation of Eqs.~\eqref{Eq:F_n_normal_ordered} and~\eqref{Eq:F_n_F_n} is largely facilitated by the symmetric notation [Eq.~\eqref{Eq:def_f_qm}] for phonon creation and annihilation operators.
The operators $f_{qm}$, which satisfy $B_q=\sum_m f_{qm}$, are analogous to the dissipaton operators of the DEOM theory~\cite{JChemPhys.140.054105,FrontPhys.11.110306,MolPhys.116.780,JChemPhys.157.170901}.
The model considered here, however, lacks explicit dissipation, so that the properties of the $f$ operators in Eq.~\eqref{Eq:def_f_qm} are somewhat different from the properties of dissipaton operators summarized in, e.g., Sec.~3.1 of Ref.~\onlinecite{MolPhys.116.780}.
Nevertheless, the generalized Wick's theorem, which is the most important ingredient of our theoretical and computational framework, turns out to assume the same form as in the DEOM theory.

The correlation functions of $f$ operators in the free-phonon ensemble are [see Eq.~\eqref{Eq:free-phonon-autocorrelation} and cf. Eq.~(3.2) of Ref.~\onlinecite{MolPhys.116.780}]
\begin{equation}
\label{Eq:dissipation_corr_functs}
\begin{split}
    \left\langle f_{2}^{(I)}(t)f_{1}^{(I)}(0)\right\rangle_\mathrm{ph}=\delta_{m_1\overline{m_2}}\eta_{q_2q_1m_2}e^{-\mu_{m_2}t},\quad
    \left\langle f_{1}^{(I)}(0)f_{2}^{(I)}(t)\right\rangle_\mathrm{ph}=\delta_{m_1\overline{m_2}}\eta_{\overline{q_2}\:\overline{q_1}\:\overline{m_2}}^*e^{-\mu_{m_2}t}.
\end{split}
\end{equation}
The $f$ operators obey $f_{qm}^\dagger=f_{\overline{q}\overline{m}}$ [cf. the text following Eq.~(3.6) of Ref.~\onlinecite{MolPhys.116.780}], and the commutation relation $[f_2,f_1]=\delta_{m_1\overline{m_2}}\delta_{q_1\overline{q_2}}(-1)^{m_2}\left(\frac{g}{\sqrt{N}}\right)^2$ [cf. the text following Eq.~(3.7) of Ref.~\onlinecite{MolPhys.116.780}].
The contraction of operators $f_2$ and $f_1$ is defined in the standard manner~\cite{Fetter-Walecka-book}
\begin{equation}
\label{Eq:def_contraction}
\begin{split}
    \wick{\c f_{2}\c f_{1}}&=f_{2}f_{1}-\normOrd{f_{2}f_{1}}\\
    &=\langle f_{2}f_{1}\rangle_\mathrm{ph}-\langle\normOrd{f_{2}f_{1}}\rangle_\mathrm{ph}.
\end{split}
\end{equation}
The second line of Eq.~\eqref{Eq:def_contraction} can be checked by direct inspection, and it emphasizes that the contraction of two operators is a c-number.
The equilibrium expectation values of a product of $f$ operators is determined by the two-point expectation value [cf. Eq.~(3.8) of Ref.~\onlinecite{MolPhys.116.780}]
\begin{equation}
\label{Eq:2-point-expt-values}
    \langle f_{2}f_{1}\rangle_\mathrm{ph}=\delta_{m_1\overline{m_2}}\eta_{q_2q_1m_2},\quad
    \langle f_{1}f_{2}\rangle_\mathrm{ph}=\delta_{m_1\overline{m_2}}\eta_{\overline{q_2}\:\overline{q_1}\:\overline{m_2}}^*,
\end{equation}
which can be obtained by letting $t\to +0$ in Eq.~\eqref{Eq:dissipation_corr_functs}.
All the properties above, and in particular Eqs.~\eqref{Eq:def_contraction} and~\eqref{Eq:2-point-expt-values}, heavily rely on the assumption of undamped phonons, for which the coefficients $\eta_{q_2q_1m}$ ($\mu_m$) are purely real (imaginary).
In a general setup with dissipation, in which $\eta_{q_2q_1m}$ ($\mu_m$) have nonzero imaginary (positive real) parts, an appropriate generalization of Eq.~\eqref{Eq:def_contraction} is under debate~\cite{su2025nonperturbativeopenquantumdynamics}.

The operators $F_\mathbf{n}^{(n)}$ in Eq.~\eqref{Eq:ADM_multiphonon_config} ultimately stem from the quantum dynamics $e^{-iH_\mathrm{tot}t}\dots e^{iH_\mathrm{tot}t}$ of the total carrier--phonon system, and as such they do not depend on the initial condition $\rho_\mathrm{tot}(0)$ from which the evolution of the interacting electron--phonon system starts.
It is, therefore, possible and most convenient to obtain $F_\mathbf{n}^{(n)}$ starting from the factorized initial condition $\rho_\mathrm{tot}(0)=\rho(0)\rho_\mathrm{ph}^\mathrm{eq}$, when we can utilize the definitions of HEOM auxiliaries in Eqs.~\eqref{Eq:define-adm-fvift} or~\eqref{Eq:define-adm-fvift-momentum-representation}.

The proof of Eq.~\eqref{Eq:F_n_normal_ordered} starts from evaluating the partial trace $\mathrm{Tr}_\mathrm{ph}\left\{B_{q_n}\dots B_{q_1}\rho_\mathrm{tot}(t)\right\}$.
To that end, we introduce auxiliary fields $\xi_q(s)$ such that the $\xi$-dependent total DM in the interaction picture reads ($\overline{\mathcal{T}}$ is the antichronological time-ordering sign)~\cite{JChemPhys.145.224105}
\begin{equation}
    \rho_{\mathrm{tot},\xi}^{(I)}(t)=\mathcal{T}\exp\left\{-i\sum_q\int_0^t ds[V_q^{(I)}(s)+\xi_q(s)]B_q^{(I)}(s)\right\}\rho(0)\rho_\mathrm{ph}^\mathrm{eq}\overline{\mathcal{T}}\exp\left\{i\sum_q\int_0^t ds\:V_q^{(I)}(s)B_q^{(I)}(s)\right\}.
\end{equation}
Then,
\begin{equation}
    \mathrm{Tr}_\mathrm{ph}\left\{B_{q_n}\dots B_{q_1}\rho_\mathrm{tot}(t)\right\}=i^n\left[\frac{\delta^n\rho_\xi(t)}{\delta\xi_{q_n}(t)\dots\delta\xi_{q_1}(t)}\right]_{\xi=0},
\end{equation}
where the $\xi$-dependent RDM in the interaction picture is
\begin{equation}
    \rho_{\xi}^{(I)}(t)=\mathrm{Tr}_\mathrm{ph}\rho_{\mathrm{tot},\xi}^{(I)}(t)=\mathcal{T}e^{-\Phi_\xi(t)}\rho(0).
\end{equation}
The $\xi$-dependent superoperator $\Phi_\xi(t)$ differs from the superoperator $\Phi(t)$ [Eq.~\eqref{Eq:def_Phi_t}] by
\begin{equation}
    \Phi_\xi(t)-\Phi(t)=i\sum_{qm}\int_0^t ds\:\xi_q(s)\varphi_{qm}(s)+\sum_{q_2q_1m}\int_0^t ds_2\int_0^{s_2}ds_1\:e^{-\mu_m(s_2-s_1)}\left[V_{q_2}^{(I)}(s_2)^\times+\xi_{q_2}(s_2)\right]\eta_{q_2q_1m}\xi_{q_1}(s_1).
\end{equation}
Only the first two functional derivatives of $\Phi_\xi(t)$ with respect to the auxiliary fields $\xi_q(s)$ are nonzero:
\begin{equation}
    \left[\frac{\delta\Phi_\xi(t)}{\delta\xi_{q_1}(t)}\right]_{\xi=0}=i\sum_{m_1}\varphi_{1}(t),
\end{equation}
\begin{equation}
\label{Eq:2nd-functional-derivative}
    \frac{\delta^2\Phi_\xi(t)}{\delta\xi_{q_2}(t)\delta\xi_{q_1}(t)}=
    \sum_{m_2m_1}\langle f_2f_1\rangle_\mathrm{ph}.
\end{equation}
In Eq.~\eqref{Eq:2nd-functional-derivative} we made use of Eq.~\eqref{Eq:2-point-expt-values}.
We eventually obtain
\begin{equation}
\label{Eq:nth-functional-derivative-interaction-picture}
\begin{split}
    \frac{\delta^n\rho_\xi^{(I)}(t)}{\delta\xi_{q_n}(t)\dots\delta\xi_{q_1}(t)}=&i^{2n}\mathcal{T}\prod_{a=1}^n\frac{\delta\Phi_\xi(t)}{\delta\xi_{q_a}(t)}e^{-\Phi_\xi(t)}\rho(0)\\
    &+i^{2n-2}\sideset{}{^n}\sum_{(ij)}\frac{\delta^2\Phi_\xi(t)}{\delta\xi_{q_j}(t)\delta\xi_{q_i}(t)}\mathcal{T}\prod_{\substack{a=1\\a\neq i,j}}^n\frac{\delta\Phi_\xi(t)}{\delta\xi_{q_a}(t)}e^{-\Phi_\xi(t)}\rho(0)\\
    &+i^{2n-4}\sideset{}{^n}\sum_{(ij)(rs)}\frac{\delta^2\Phi_\xi(t)}{\delta\xi_{q_s}(t)\delta\xi_{q_r}(t)}\frac{\delta^2\Phi_\xi(t)}{\delta\xi_{q_j}(t)\delta\xi_{q_i}(t)}\mathcal{T}\prod_{\substack{a=1\\a\neq i,j,k,l}}^n\frac{\delta\Phi_\xi(t)}{\delta\xi_{q_a}(t)}e^{-\Phi_\xi(t)}\rho(0)\\
    &+\dots
\end{split}
\end{equation}
Setting $\xi=0$ in Eq.~\eqref{Eq:nth-functional-derivative-interaction-picture},
remembering the definition of the auxiliary operators [Eq.~\eqref{Eq:define-adm-fvift-momentum-representation}],
and transferring to the Schr\"{o}dinger picture, we find
\begin{equation}
\label{Eq:prod_Bs_in_rhos_final}
\begin{split}
    &\mathrm{Tr}_\mathrm{ph}\left\{B_{q_n}\dots B_{q_1}\rho_\mathrm{tot}(t)\right\}=\sum_{m_n\dots m_1}\rho_\mathbf{n}^{(n)}(t)\\
    &+\sum_{m_n\dots m_1}\sideset{}{^n}\sum_{(ij)}\langle f_{j}f_{i}\rangle_\mathrm{ph}\rho_{\mathbf{n}_{ji}^-}^{(n-2)}(t)\\
    &+\sum_{m_n\dots m_1}\:\sideset{}{^n}\sum_{(ij)(rs)}\langle f_{s}f_{r}\rangle_\mathrm{ph}\langle f_{j}f_{i}\rangle_\mathrm{ph}\rho_{\mathbf{n}_{srji}^{-}}^{(n-4)}(t)\\
    &+\dots
\end{split}
\end{equation}
The final term on the RHS of Eq.~\eqref{Eq:prod_Bs_in_rhos_final} is proportional to the RDM if $n$ is even, while it is a linear combination of the first-level auxiliary operators for odd $n$.
Upon inserting 
$B_q=\sum_m f_{qm}$ and Eq.~\eqref{Eq:ADM_multiphonon_config} into Eq.~\eqref{Eq:prod_Bs_in_rhos_final}, we obtain
\begin{equation}
\label{Eq:before_first_F_in_f}
\begin{split}
    \sum_{m_n\dots m_1}\left[f_{n}\dots f_{1}-F_\mathbf{n}^{(n)}-\sideset{}{^n}\sum_{(ij)}\langle f_{j}f_{i}\rangle_\mathrm{ph}F_{\mathbf{n}_{ji}^-}^{(n-2)}\right. \\ \left. -\sideset{}{^n}\sum_{(ij)(rs)}\langle f_{s}f_{r}\rangle_\mathrm{ph}\langle f_{j}f_{i}\rangle_\mathrm{ph}F_{\mathbf{n}_{srji}^{-}}^{(n-4)}-\dots\right]=0.
\end{split}
\end{equation}
Each term in the square brackets of Eq.~\eqref{Eq:before_first_F_in_f} should be separately equal to zero.
The situation is, however, complicated by the fact that different terms behave differently under permutations of pairs $(q_i,m_i)$.
Equation~\eqref{Eq:prod_Bs_in_rhos_final} is invariant under permutations of momenta $q_i$ because (i) the operators $B_q$ mutually commute, (ii) the auxiliaries are invariant under permutations of the involved momenta (the dummy indices $m_i$ can be permuted at will) and (iii) the expectation value $\langle f_{j}f_{i}\rangle_\mathrm{ph}$ is invariant under permutation $q_j\leftrightarrow q_i$ [see Eqs.~\eqref{Eq:def_c_0},~\eqref{Eq:def_c_1}, and~\eqref{Eq:2-point-expt-values}].
On the other hand, the expression in the square brackets of Eq.~\eqref{Eq:before_first_F_in_f} is not invariant under permutations of pairs $(q_i,m_i)$ because the operators $f_i$ do not commute. 
To make the product $f_{n}\dots f_{1}$ invariant under permutations of indices $(q_i,m_i)$, we resort to Wick's theorem in its operator form~\cite{Fetter-Walecka-book}:
\begin{equation}
\label{Eq:Wick_operatorial}
\begin{split}
    f_{n}\dots f_{1}=&\normOrd{\prod_{a=1}^n f_a}+\sideset{}{^n}\sum_{(ij)}\wick{\c f_{j}\c f_{i}}\normOrd{\prod_{\substack{a=1\\a\neq i,j}}^n f_{a}}\\
    &+\sideset{}{^n}\sum_{(ij)(rs)}\wick{\c f_{s}\c f_{r}}\wick{\c f_{j}\c f_{i}}\normOrd{\prod_{\substack{a=1\\a\neq i,j,r,s}}^n f_{a}}\\
    &+\dots
\end{split}
\end{equation}
Combining Eqs.~\eqref{Eq:before_first_F_in_f} and~\eqref{Eq:Wick_operatorial}, we obtain
\begin{equation}
\label{Eq:first_F_in_f}
\begin{split}
    &F_\mathbf{n}^{(n)}=\normOrd{\prod_{a=1}^n f_a}
    +\sideset{}{^n}\sum_{(ij)}\left[\wick{\c f_j\c f_i}\normOrd{\prod_{\substack{a=1\\a\neq i,j}}^n f_a}-\langle f_jf_i\rangle_\mathrm{ph}F_{\mathbf{n}_{ji}^-}^{(n-2)}\right]\\
    &+\sideset{}{^n}\sum_{(ij)(rs)}\left[\wick{\c f_s\c f_r}\wick{\c f_j\c f_i}\normOrd{\prod_{\substack{a=1\\a\neq i,j,r,s}}^n f_a}-\langle f_{s}f_{r}\rangle_\mathrm{ph}\langle f_{j}f_{i}\rangle_\mathrm{ph}F_{\mathbf{n}_{srji}^{-}}^{(n-4)}\right]+\dots
\end{split}
\end{equation}
To express $F_\mathbf{n}^{(n)}$ in terms of $f_{qm}$ only, we have to recursively insert analogues of Eq.~\eqref{Eq:first_F_in_f} for lower-order phonon operators $F_{\mathbf{n}_{ji}^-}^{(n-2)}$, $F_{\mathbf{n}_{srji}^{-}}^{(n-4)}$, etc., into Eq.~\eqref{Eq:first_F_in_f} itself.
This is done order by order in phonon assistance.
We illustrate the procedure on the example of $n$-, $(n-2)$-, and $(n-4)$-phonon contributions to $F_\mathbf{n}^{(n)}$, for which in Eq.~\eqref{Eq:first_F_in_f} we replace
\begin{equation}
    F_{\mathbf{n}_{srji}^{-}}^{(n-4)}\to\normOrd{\prod_{\substack{a=1\\a\neq i,j,r,s}}^nf_{a}}
\end{equation}
and
\begin{equation}
\begin{split}
    F_{\mathbf{n}_{ji}^-}^{(n-2)}\to &\normOrd{\prod_{\substack{a=1\\a\neq i,j}}^n f_{a}}-\sideset{}{^{n-2}}\sum_{(rs)}\langle\normOrd{f_{s}f_{r}}\rangle_\mathrm{ph}\normOrd{\prod_{\substack{a=1\\a\neq i,j,r,s}}^n f_{a}}.
\end{split}
\end{equation}
In the resulting equation, when grouping terms containing the same number of phonons, we observe the following formal replacement $\displaystyle{\sideset{}{^n}\sum_{(ij)}\sideset{}{^{n-2}}\sum_{(rs)}=2\sideset{}{^n}\sum_{(ij)(rs)}}$, which reflects the fact that the order of pairs in immaterial when these are chosen out of $n$ elements from the outset.
We eventually obtain Eq.~\eqref{Eq:F_n_normal_ordered}.

The derivation of Eq.~\eqref{Eq:F_n_F_n} is also performed recursively.
Equation~\eqref{Eq:F_n_normal_ordered}, which can be recast as
\begin{equation}
\label{Eq:F_n_normal_ordered_inverted}
\begin{split}
    &\normOrd{\prod_{a=1}^n f_{a}}=F_\mathbf{n}^{(n)}\\&+\sideset{}{^n}\sum_{(ij)}\langle\normOrd{f_{j}f_{i}}\rangle_\mathrm{ph}\normOrd{\prod_{\substack{a=1\\a\neq i,j}}^n f_{a}}\\
    &-\sideset{}{^n}\sum_{(ij)(rs)}\langle\normOrd{f_{s}f_{r}}\rangle_\mathrm{ph}\langle\normOrd{f_{j}f_{i}}\rangle_\mathrm{ph}\normOrd{\prod_{\substack{a=1\\a\neq i,j,r,s}}^n f_{a}}\\
    &+\dots
\end{split}
\end{equation}
is recursively inserted into itself to express the normal-order product $\normOrd{\prod_{a=1}^n f_{a}}$ of phonon operators in terms of operators describing irreducible phonon correlations.
Here, we concentrate on deriving the contributions of irreducible correlations comprising $(n-2)$ ($F_{\mathbf{n}_{ji}^-}^{(n-2)}$) and $(n-4)$ ($F_{\mathbf{n}_{srji}^-}^{(n-4)}$) phonons to $n$-phonon irreducible correlations embodied in $F_\mathbf{n}^{(n)}$.
To that end, using Eq.~\eqref{Eq:F_n_normal_ordered_inverted}, we insert
\begin{equation}
    \normOrd{\prod_{\substack{a=1\\a\neq i,j,r,s}}^n f_{a}}\to F_{\mathbf{n}_{srji}^-}^{(n-4)},\quad
    \normOrd{\prod_{\substack{a=1\\a\neq i,j}}^n f_{a}}\to F_{\mathbf{n}_{ji}^-}^{(n-2)}+\sideset{}{^{(n-2)}}\sum_{(rs)}\langle\normOrd{f_sf_r}\rangle_\mathrm{ph}F_{\mathbf{n}_{srji}^-}^{(n-4)}
\end{equation}
into Eq.~\eqref{Eq:F_n_normal_ordered}, and consider the formal replacement $\displaystyle{\sideset{}{^n}\sum_{(ij)}\sideset{}{^{n-2}}\sum_{(rs)}=2\sideset{}{^n}\sum_{(ij)(rs)}}$ to obtain the three terms on the RHS of Eq.~\eqref{Eq:F_n_F_n}.

\section{Proof of the generalized Wick's theorem}
\label{App:GWT_rigorously_proven}

The crux of the proof of Eq.~\eqref{Eq:GWT_right} [Eq.~\eqref{Eq:GWT_left} is proven analogously] is the rule by which an operator is introduced into a normally ordered string of operators~\cite{Fetter-Walecka-book}:
\begin{equation}
\label{Eq:rule_normal_ordered_right}
\begin{split}
\normOrd{\prod_{a=1}^n f_{a}}f_{{n+1}}=\normOrd{\prod_{a=1}^{n+1}f_{a}}+\sum_{i=1}^n\wick{\c f_{i} \c f_{{n+1}}}\normOrd{\prod_{\substack{a=1\\a\neq i}}^n f_a}.
\end{split}
\end{equation}
We use Eq.~\eqref{Eq:rule_normal_ordered_right} to express $F_\mathbf{n}^{(n)}f_{{n+1}}$ in terms of normally ordered products:
\begin{equation}
\label{Eq:first_in_B}
\begin{split}
    &F_\mathbf{n}^{(n)}f_{{n+1}}=\normOrd{\prod_{a=1}^{n+1}f_{a}}+\sum_{i=1}^n\wick{\c f_{i} \c f_{{n+1}}}\normOrd{\prod_{\substack{a=1\\a\neq i}}^nf_{a}}\\
    &-\sideset{}{^n}\sum_{(ij)}\langle\normOrd{f_{j}f_{i}}\rangle_\mathrm{ph}\left[\normOrd{\prod_{\substack{a=1\\a\neq i,j}}^{n+1}f_{a}}+\sum_{\substack{r=1\\r\neq i,j}}^n\wick{\c f_{r} \c f_{{n+1}}}\normOrd{\prod_{\substack{a=1\\a\neq i,j,r}}^nf_{a}}\right]\\
    &+\sideset{}{^n}\sum_{(ij)(rs)}\langle\normOrd{f_{s}f_{r}}\rangle_\mathrm{ph}\langle\normOrd{f_{j}f_{i}}\rangle_\mathrm{ph}\left[\normOrd{\prod_{\substack{a=1\\a\neq i,j,r,s}}^{n+1}f_{a}}+\sum_{\substack{v=1\\v\neq i,j,r,s}}^n\wick{\c f_{v} \c f_{{n+1}}}\normOrd{\prod_{\substack{a=1\\a\neq i,j,r,s,v}}^n f_{a}}\right]\\
    &-\dots
\end{split}
\end{equation}
We proceed by grouping the terms on the RHS of Eq.~\eqref{Eq:first_in_B} based on the number of phonon operators that do not participate in expectation values, i.e.,
\begin{equation}
\label{Eq:B2}
    F_\mathbf{n}^{(n)}f_{{n+1}}=\left[F_\mathbf{n}^{(n)}f_{{n+1}}\right]_{n+1}+\left[F_\mathbf{n}^{(n)}f_{{n+1}}\right]_{n-1}+\left[F_\mathbf{n}^{(n)}f_{{n+1}}\right]_{n-3}+\dots
\end{equation}
The only term containing $n+1$ phonon operators outside of expectation values is the first term on the RHS, i.e.,
\begin{equation}
    \left[F_\mathbf{n}^{(n)}f_{{n+1}}\right]_{n+1}=\normOrd{\prod_{a=1}^{n+1}f_{a}}
\end{equation}
The terms containing $n-1$ phonon operators are
\begin{equation}
\label{Eq:first_in_B_n-1_ph_ops}
\begin{split}
    \left[F_\mathbf{n}^{(n)}f_{{n+1}}\right]_{n-1}&=\sum_{i=1}^n\langle f_{i}f_{{n+1}}\rangle_\mathrm{ph}\normOrd{\prod_{\substack{a=1\\a\neq i}}^nf_{a}}\\
    &-\sum_{i=1}^n\langle\normOrd{f_{i}f_{{n+1}}}\rangle_\mathrm{ph}\normOrd{\prod_{\substack{a=1\\a\neq i}}^nf_{a}}
    -\sideset{}{^n}\sum_{(ij)}\langle\normOrd{f_{j}f_{i}}\rangle_\mathrm{ph}\normOrd{\prod_{\substack{a=1\\a\neq i,j}}^{n+1}f_{a}}\\
    &=\sum_{i=1}^n\langle f_{i}f_{{n+1}}\rangle_\mathrm{ph}\normOrd{\prod_{\substack{a=1\\a\neq i}}^nf_{a}}-\sideset{}{^{n+1}}\sum_{(ij)}\langle\normOrd{f_{j}f_{i}}\rangle_\mathrm{ph}\normOrd{\prod_{\substack{a=1\\a\neq i,j}}^{n+1}f_{a}}
\end{split}
\end{equation}
In going from the first to the second equality of Eq.~\eqref{Eq:first_in_B_n-1_ph_ops}, we observed that all possible two-combinations from a set of $n+1$ elements $\{1,\dots,n+1\}$ can be obtained from all possible two-combinations from a set of $n$ elements $\{1,\dots,n\}$ by adding the $n$ missing pairs $\{(n+1,1),\dots,(n+1,n)\}$.
The terms containing $n-3$ phonon operators read
\begin{equation}
\label{Eq:first_in_B_n-3_ph_ops}
\begin{split}
    \left[F_\mathbf{n}^{(n)}f_{{n+1}}\right]_{n-3}&=-\sideset{}{^n}\sum_{(ij)}\sum_{\substack{r=1\\r\neq i,j}}^n\langle\normOrd{f_{j}f_{i}}\rangle_\mathrm{ph}\langle f_{r}f_{{n+1}}\rangle_\mathrm{ph}\normOrd{\prod_{\substack{a=1\\a\neq i,j,r}}^n f_{a}}\\
    &+\sideset{}{^n}\sum_{(ij)}\sum_{\substack{r=1\\r\neq i,j}}^n\langle\normOrd{f_{j}f_{i}}\rangle_\mathrm{ph}\langle\normOrd{f_{r}f_{{n+1}}}\rangle_\mathrm{ph}\normOrd{\prod_{\substack{a=1\\a\neq i,j,r}}^n f_{a}}\\
    &+\sideset{}{^n}\sum_{(ij)(rs)}\langle\normOrd{f_{s}f_{r}}\rangle_\mathrm{ph}\langle\normOrd{f_{j}f_{i}}\rangle_\mathrm{ph}\normOrd{\prod_{\substack{a=1\\a\neq i,j,r,s}}^{n+1}f_{a}}
\end{split}
\end{equation}
The first term on the RHS of Eq.~\eqref{Eq:first_in_B_n-3_ph_ops} contains $\binom{n}{2}(n-2)=n\binom{n-1}{2}$ summands, and exchanging the order of summations we recast it as
\begin{equation}
    -\sum_{i=1}^n\langle f_{i}f_{{n+1}}\rangle_\mathrm{ph}\sideset{}{^{n-1}}\sum_{\substack{(jr)\\j,r\neq i}}\langle\normOrd{f_{r}f_{j}}\rangle_\mathrm{ph}\normOrd{\prod_{\substack{a=1\\a\neq i,j,r}}^n f_{a}}.
\end{equation}
The other two terms on the RHS of Eq.~\eqref{Eq:first_in_B_n-3_ph_ops} contain $\binom{n}{2}(n-2)+\frac{1}{2!}\binom{n}{2}\binom{n-2}{2}=\frac{1}{2!}\binom{n+1}{2}\binom{n-1}{2}$ summands in total, and these can be regrouped as
\begin{equation}
    \sideset{}{^{n+1}}\sum_{\substack{(ij)(rs)}}\langle\normOrd{f_{s}f_{r}}\rangle_\mathrm{ph}\langle\normOrd{f_{j}f_{i}}\rangle_\mathrm{ph}\normOrd{\prod_{\substack{a=1\\a\neq i,j,r,s}}^{n+1}f_{a}}
\end{equation}
We finally obtain
\begin{equation}
\label{Eq:B_ultimate}
\begin{split}
    &F_\mathbf{n}^{(n)}f_{{n+1}}=\\&\normOrd{\prod_{a=1}^{n+1}f_{a}}-\sideset{}{^{n+1}}\sum_{(ij)}\langle\normOrd{f_{j}f_{i}}\rangle_\mathrm{ph}\normOrd{\prod_{\substack{a=1\\a\neq i,j}}^{n+1}f_{a}}+\sideset{}{^{n+1}}\sum_{\substack{(ij)(rs)}}\langle\normOrd{f_{s}f_{r}}\rangle_\mathrm{ph}\langle\normOrd{f_{j}f_{i}}\rangle_\mathrm{ph}\normOrd{\prod_{\substack{a=1\\a\neq i,j,r,s}}^{n+1}f_{a}}-\dots\\&+\sum_{i=1}^n\langle f_{i}f_{{n+1}}\rangle_\mathrm{ph}\left[\normOrd{\prod_{\substack{a=1\\a\neq i}}^nf_{a}}-\sideset{}{^{n-1}}\sum_{\substack{(jr)\\j,r\neq i}}\langle\normOrd{f_{r}f_{j}}\rangle_\mathrm{ph}\normOrd{\prod_{\substack{a=1\\a\neq i,j,r}}^n f_{a}}+\dots\right]
\end{split}
\end{equation}
Using Eq.~\eqref{Eq:F_n_normal_ordered}, the three terms in the first line of the RHS of Eq.~\eqref{Eq:B_ultimate} can be recognized as the leading three terms (with respect to the number of phonons) of $F_{\mathbf{n}_{{n+1}}^+}^{(n+1)}$.
Similarly, the two terms within the square brackets in the second line of the RHS of Eq.~\eqref{Eq:B_ultimate} can be recognized as the leading two terms of $F_{\mathbf{n}_{i}^-}^{(n-1)}$.
Since the remaining terms, containing an even smaller number of phonons, can be obtained by considering further terms in Eq.~\eqref{Eq:B2}, the proof of the generalized Wick's theorem can be considered completed.

\section{Derivation of HEOM closing schemes}
\label{App:hierarchy_closing}
Let us assume that vector $\mathbf{D}$ of non-negative integers $D_{qm}$ ($q\neq 0;m=0,1$) is such that $\sum_{qm}D_{qm}=D$, where $D$ is the maximum hierarchy depth.
The term that couples the auxiliary $\rho_\mathbf{D}^{(D)}(t)$ with the auxiliaries at depth $D+1$ is
\begin{equation}
\label{Eq:coupling_D_to_D_plus_1}
\begin{split}
    [\partial_t\rho_\mathbf{D}^{(D)}(t)]_\mathrm{close}=&-i\sum_{qm}\sqrt{1+D_{qm}}\sqrt{|c_m|}V_q^\times\rho_{\mathbf{D}_{qm}^+}^{(D+1)}(t).
\end{split}
\end{equation}
The evolution of $\rho_{\mathbf{D}_{qm}^+}^{(D+1)}(t)$ is governed by (with $\mathcal{L}_\mathrm{e}O=[H_\mathrm{e},O],V^>O=VO,V^<O=OV$)
\begin{equation}
\label{Eq:eom_D_plus_1}
\begin{split}
    \partial_t\rho_{\mathbf{D}_{qm}^+}^{(D+1)}(t)=&-(i\mathcal{L}_\mathrm{e}+\mu_\mathbf{D}+\mu_m)\rho_{\mathbf{D}_{qm}^+}^{(D+1)}(t)\\
    &-i\sum_{q'm'}\sqrt{1+D_{q'm'}+\delta_{q'q}\delta_{m'm}}\sqrt{|c_{m'}|}V_{q'}^\times\rho_{\mathbf{D}_{qm,q'm'}^{+,+}}^{(D+2)}(t)\\
    &-i\sum_{q'm'}\frac{\sqrt{D_{q'm'}+\delta_{q'q}\delta_{m'm}}}{\sqrt{|c_{m'}|}}\left[c_{m'}V_{\overline{q'}}^>-c_{\overline{m'}}^*V_{\overline{q'}}^<\right]\rho_{\mathbf{D}_{qm,q'm'}^{+,-}}^{(D)}(t).
\end{split}
\end{equation}
\subsection{Markovian and adiabatic scheme}
\label{SApp:MA_closing}
We developed and tested this scheme on the one-dimensional Holstein model in Ref.~\onlinecite{JChemPhys.159.094113}.
It transforms Eq.~\eqref{Eq:eom_D_plus_1} by neglecting the hierarchical couplings to auxiliaries at depth $D+2$ [the second term on the RHS of Eq.~\eqref{Eq:eom_D_plus_1}] and retaining only the coupling to $\rho_\mathbf{D}^{(D)}(t)$ for which Eq.~\eqref{Eq:coupling_D_to_D_plus_1} is formulated [the summand with $q'=q$ and $m'=m$ in the third term on the RHS of Eq.~\eqref{Eq:eom_D_plus_1}].
As $\rho_{\mathbf{D}_{qm}^+}^{(D+1)}(0)=0$ (both the imaginary-time and real-time HEOM are truncated at maximum depth $D$), the solution of the transformed Eq.~\eqref{Eq:eom_D_plus_1} reads
\begin{equation}
\label{Eq:D_plus_1_solution_1}
    \rho_{\mathbf{D}_{qm}^+}^{(D+1)}(t)=-i\frac{\sqrt{1+D_{qm}}}{\sqrt{|c_{m}|}}\int_0^t ds\:e^{-\mu_m s}\:e^{-i\mathcal{L}_\mathrm{e} s}[c_mV_{\overline{q}}^>-c_{\overline{m}}^*V_{\overline{q}}^<]e^{i\mathcal{L}_\mathrm{e} s}\:e^{-(i\mathcal{L}_\mathrm{e}+\mu_\mathbf{D})t}\widetilde{\rho}_\mathbf{D}^{(D)}(t-s).
\end{equation}
In Eq.~\eqref{Eq:D_plus_1_solution_1}, $\widetilde{\rho}_\mathbf{D}^{(D)}(s)=e^{(i\mathcal{L}_\mathrm{e}+\mu_\mathbf{D})s}\rho_\mathbf{D}^{(D)}(s)$ denotes the slowly changing part of the auxiliary $\rho_\mathbf{D}^{(D)}(s)$.
The Markovian approximation replaces $\widetilde{\rho}_\mathbf{D}^{(D)}(t-s)\approx\widetilde{\rho}_\mathbf{D}^{(D)}(t)$ in Eq.~\eqref{Eq:D_plus_1_solution_1}.
The adiabatic approximation extends the upper integration limit in Eq.~\eqref{Eq:D_plus_1_solution_1} to infinity.
Physically, the final, $(D+1)$-st single phonon-assisted process, is assumed to be temporally well separated from the $D$ single phonon-assisted processes that are treated exactly.
Combining Markovian and adiabatic approximations, we express the auxiliaries at depth $D+1$ in terms of $\rho_\mathbf{D}^{(D)}(t)$ as follows:
\begin{equation}
\label{Eq:D_plus_1_Markov_adiabatic}
    \rho_{\mathbf{D}_{qm}^+}^{(D+1)}(t)=-i\frac{\sqrt{1+D_{qm}}}{\sqrt{|c_{m}|}}\int_0^{+\infty} ds\:e^{-\mu_m s}\left[c_m V_{\overline{q}}^{(I)}(-s)\rho_\mathbf{D}^{(D)}(t)-c_{\overline{m}}^*\rho_\mathbf{D}^{(D)}(t)V_{\overline{q}}^{(I)}(-s)\right].
\end{equation}
Inserting Eq.~\eqref{Eq:D_plus_1_Markov_adiabatic} into Eq.~\eqref{Eq:coupling_D_to_D_plus_1}, we obtain the following closing term:
\begin{equation}
\begin{split}
    [\partial_t\rho_\mathbf{D}^{(D)}(t)]_\mathrm{close}=&-\sum_{qm}(1+D_{qm})\int_0^{+\infty}ds\:e^{-\mu_m s}\left[c_mV_qV_{\overline{q}}^{(I)}(-s)\rho_\mathbf{D}^{(D)}(t)+c_{\overline{m}}^*\rho_\mathbf{D}^{(D)}(t)V_{\overline{q}}^{(I)}(-s)V_q\right]\\
    &+\sum_{qm}(1+D_{qm})\int_0^{+\infty}ds\:e^{-\mu_m s}\left[c_mV_{\overline{q}}^{(I)}(-s)\rho_\mathbf{D}^{(D)}(t)V_q+c_{\overline{m}}^*V_q\rho_\mathbf{D}^{(D)}(t)V_{\overline{q}}^{(I)}(-s)\right].
\end{split}
\end{equation}
The matrix element $\langle k|\dots|k+k_\mathbf{D}\rangle$ of the last equation reads
\begin{equation}
\label{Eq:closing_before_RPA}
\begin{split}
&[\partial_t\langle k|\rho_\mathbf{D}^{(D)}(t)|k+k_\mathbf{D}\rangle]_\mathrm{close}=\\&-\left[\sum_{qm}(1+D_{qm})c_m |M(k-q,q)|^2\int_0^{+\infty}ds\:e^{-[\mu_m+i(\varepsilon_{k-q}-\varepsilon_k)]s}\right]\langle k|\rho_\mathbf{D}^{(D)}(t)|k+k_\mathbf{D}\rangle\\&-\left[\sum_{qm}(1+D_{qm})c_{\overline{m}}^*|M(k+k_\mathbf{D},q)|^2\int_0^{+\infty}ds\:e^{-[\mu_m+i(\varepsilon_{k+k_\mathbf{D}}-\varepsilon_{k+k_\mathbf{D}+q})]s}\right]\langle k|\rho_\mathbf{D}^{(D)}(t)|k+k_\mathbf{n}\rangle\\
&+\sum_{qm}(1+D_{qm})c_m M(k,q)^*M(k+k_\mathbf{D},q)\int_0^{+\infty}ds\:e^{-[\mu_m+i(\varepsilon_{k}-\varepsilon_{k+q})]s}\langle k+q|\rho_\mathbf{D}^{(D)}(t)|k+k_\mathbf{D}+q\rangle\\
&+\sum_{qm}(1+D_{qm})c_{\overline{m}}^*M(k,q)^*M(k+k_\mathbf{D},q)\int_0^{+\infty}ds\:e^{-[\mu_m+i(\varepsilon_{k+k_\mathbf{D}+q}-\varepsilon_{k+k_\mathbf{D}})]s}\langle k+q|\rho_\mathbf{D}^{(D)}(t)|k+k_\mathbf{D}+q\rangle.
\end{split}
\end{equation}
The third and the fourth term on the RHS of Eq.~\eqref{Eq:closing_before_RPA} involve summations of complex-valued $q$-dependent quantities $M(k,q)^*M(k+k_\mathbf{D},q)\langle k+q|\rho_\mathbf{D}^{(D)}(t)|k+k_\mathbf{D}+q\rangle$.
In the random-phase approximation~\cite{kuhncontribution,RevModPhys.74.895,PhysRevB.92.235208}, these terms are considered as vanishing.
In the first two terms on the RHS of Eq.~\eqref{Eq:closing_before_RPA}, which contain $q$-independent matrix elements of $\rho_\mathbf{D}^{(D)}(t)$, we replace $1+D_{qm}\to 1$, approximate $\int_0^{+\infty}ds\:e^{-i\Omega s}\approx\pi\delta(\Omega)$ [i.e., we neglect the imaginary part that would change the free-oscillation frequency of $\langle k|\rho_\mathbf{D}^{(D)}(t)|k+k_\mathbf{D}\rangle$~\cite{kuhncontribution,RevModPhys.74.895,PhysRevB.92.235208}], and evaluate the sums over $q$ in the infinite-chain limit $N\to+\infty$.
As a result, we finally obtain Eq.~\eqref{Eq:MA_closing}.

\subsection{Derivative-resum scheme}
\label{SApp:DR_closing}
In Eq.~\eqref{Eq:eom_D_plus_1}, we assume that~\cite{JChemPhys.142.104112}
\begin{equation}
    \partial_t\rho_{\mathbf{D}_{qm}^+}^{(D+1)}(t)\approx 
    -i\sum_{q'm'}\sqrt{1+D_{q'm'}+\delta_{q'q}\delta_{m'm}}\sqrt{|c_{m'}|}V_{q'}^\times\rho_{\mathbf{D}_{qm,q'm'}^{+,+}}^{(D+2)}(t),
\end{equation}
and retain only the coupling to $\rho_\mathbf{D}^{(D)}(t)$ [the summand with $q'=q$ and $m'=m$ in the third term on the RHS of Eq.~\eqref{Eq:eom_D_plus_1}]~\cite{JChemPhys.157.054108}.
We thus obtain
\begin{equation}
\label{Eq:D_plus_1_derivative_resum}
    \rho_{\mathbf{D}_{qm}^+}^{(D+1)}(t)=-\frac{\sqrt{1+D_{qm}}}{\sqrt{|c_m|}}[\mathcal{L}_\mathrm{e}-i(\mu_\mathbf{D}+\mu_m)]^{-1}(c_mV_{\overline{q}}^>-c_{\overline{m}}^*V_{\overline{q}}^<)\rho_\mathbf{D}^{(D)}(t).
\end{equation}
Inserting Eq.~\eqref{Eq:D_plus_1_derivative_resum} into Eq.~\eqref{Eq:coupling_D_to_D_plus_1} yields the following closing term:
\begin{equation}
\label{Eq:derivative_resum_initial}
    [\partial_t\rho_\mathbf{D}^{(D)}(t)]_\mathrm{close}=i\sum_{qm}(1+D_{qm})V_q^\times[\mathcal{L}_\mathrm{e}-i(\mu_\mathbf{D}+\mu_m)]^{-1}(c_mV_{\overline{q}}^>-c_{\overline{m}}^*V_{\overline{q}}^<)\rho_\mathbf{D}^{(D)}(t).
\end{equation}
Similarly to the Markovian and adiabatic closing, we replace $1+D_{qm}\to 1$ in Eq.~\eqref{Eq:derivative_resum_initial}.
The random-phase approximation~\cite{kuhncontribution,RevModPhys.74.895,PhysRevB.92.235208} neglects the two terms in which $V_q$ and $V_{\overline{q}}$ act on $\rho_\mathbf{D}^{(D)}(t)$ from opposite sides.
In the remaining two terms, in which $V_q$ and $V_{\overline{q}}$ act on $\rho_\mathbf{D}^{(D)}(t)$ from the same side, we consider that $\mu_m$ has a small positive real part [the exponential decomposition in Eq.~\eqref{Eq:free-phonon-autocorrelation} is considered for $t>0$] and approximate $[\Delta\varepsilon-i(\mu_\mathbf{D}+\mu_m)-i0_+]^{-1}\approx i\pi\delta[\Delta\varepsilon-i(\mu_\mathbf{D}+\mu_m)]$, i.e., we neglect the imaginary part that would change the free-oscillation frequency of $\langle k|\rho_\mathbf{D}^{(D)}(t)|k+k_\mathbf{D}\rangle$.
We then evaluate the sums over $q$ in the $N\to+\infty$ limit, and finally obtain 
\begin{equation}
\label{Eq:DR_closing_final}
\begin{split}
    &\Gamma_\mathrm{DR}(k,\mathbf{D})=\frac{2g^2}{J(e^{\beta\omega_0}-1)}\left[\frac{2-\left(\frac{\varepsilon_k}{2J}\right)^2-\left(\frac{\varepsilon_{k+k_\mathbf{D}}-(N_\mathbf{D}-1)\omega_0}{2J}\right)^2}{\sqrt{1-\left(\frac{\varepsilon_{k+k_\mathbf{D}}-(N_\mathbf{D}-1)\omega_0}{2J}\right)^2}}+\frac{2-\left(\frac{\varepsilon_{k+k_\mathbf{D}}}{2J}\right)^2-\left(\frac{\varepsilon_{k}+(N_\mathbf{D}+1)\omega_0}{2J}\right)^2}{\sqrt{1-\left(\frac{\varepsilon_{k}+(N_\mathbf{D}+1)\omega_0}{2J}\right)^2}}\right]\\
    &+\frac{2g^2}{J(1-e^{-\beta\omega_0})}\left[\frac{2-\left(\frac{\varepsilon_k}{2J}\right)^2-\left(\frac{\varepsilon_{k+k_\mathbf{D}}-(N_\mathbf{D}+1)\omega_0}{2J}\right)^2}{\sqrt{1-\left(\frac{\varepsilon_{k+k_\mathbf{D}}-(N_\mathbf{D}+1)\omega_0}{2J}\right)^2}}+\frac{2-\left(\frac{\varepsilon_{k+k_\mathbf{D}}}{2J}\right)^2-\left(\frac{\varepsilon_{k}+(N_\mathbf{D}-1)\omega_0}{2J}\right)^2}{\sqrt{1-\left(\frac{\varepsilon_{k}+(N_\mathbf{D}-1)\omega_0}{2J}\right)^2}}\right],
\end{split}
\end{equation}
while $N_\mathbf{D}=-i\frac{\mu_\mathbf{D}}{\omega_0}=\sum_{q}(D_{q0}-D_{q1})$ is the net number of exchanged phonons.

In contrast to the MA scheme, which is physically motivated and considers only the final, $(D+1)$-st single phonon-assisted process, the DR scheme is less physically transparent and effectively considers all $D+1$ single phonon-assisted processes.

\section{Electron mobility in the weak-interaction limit: Predictions based on the Boltzmann equation}
\label{App:Boltzmann}
We summarize the procedure to compute carrier mobility in the weak-interaction limit using the Boltzmann (semiclassical) description of transport.

Quite generally, when the interacting carrier--phonon system is placed in an external electric field $E$, the stationary population $p_k$ of the free-carrier state $|k\rangle$ satisfies
\begin{equation}
\label{Eq:Bltz_most_general}
    E\frac{\partial p_k}{\partial k}=\left(\frac{\partial p_k}{\partial t}\right)_\mathrm{e-ph}.
\end{equation}
The collision integral $\left(\frac{\partial p_k}{\partial t}\right)_\mathrm{e-ph}$ describes the redistribution of populations due to the carrier--phonon scattering.
In the second order in the carrier--phonon interaction, one obtains (see Sec.~SV of the Supplemental Material~\cite{comment241224})
\begin{equation}
    \left(\frac{\partial p_k}{\partial t}\right)_\mathrm{e-ph}=-\sum_q w_{k+q,k}p_k+\sum_q w_{k,k+q}p_{k+q},
\end{equation}
where the transition rate from state $|k\rangle$ to state $|k+q\rangle$ reads as
\begin{equation}
\label{Eq:w_k-plus-q_k_2nd}
    w_{k+q,k}=2\pi\frac{g^2}{N}|M(k,q)|^2\sum_{\pm}\left(n_\mathrm{ph}+\frac{1}{2}\pm\frac{1}{2}\right)\delta(\varepsilon_{k+q}-\varepsilon_k\pm\omega_0).
\end{equation}
The transition rates satisfy the detailed-balance condition
\begin{equation}
 w_{k+q,k}p_{k,0}=w_{k,k+q}p_{k+q,0},   
\end{equation}
where $p_{k,0}\propto e^{-\beta\varepsilon_k}$ [we abbreviate $p_{k,0}=p_k(E=0)$] are the stationary populations of free-carrier states for $E=0$ and in the limit of weak carrier--phonon scattering.

Assuming that the external electric field is weak, and inserting $p_k\approx p_{k,0}+(\partial_E p_k)_0E$ [we abbreviate $(\partial_E p_k)_0=(\partial p_k(E)/\partial E)_{E=0}$] into Eq.~\eqref{Eq:Bltz_most_general}, we obtain the following linearized version of the Boltzmann equation [see, e.g., Eq.~(40) of Ref.~\onlinecite{RepProgPhys.83.036501}]:
\begin{equation}
\label{Eq:linearized_bltz}
    v_k\frac{\partial p_{k,0}}{\partial\varepsilon_k}=-\sum_q w_{k+q,k}(\partial_Ep_k)_0+\sum_q w_{k,k+q}(\partial_Ep_{k+q})_0.
\end{equation}
The solution for the linear-response coefficients $(\partial_E p_k)_0$ is sought in the form~\cite{RepProgPhys.83.036501}
\begin{equation}
\label{Eq:linearized_bltz_ansatz}
    (\partial_E p_k)_0=-v_k\frac{\partial p_{k,0}}{\partial\varepsilon_k}\widetilde{\tau}_k,
\end{equation}
where $k$-dependent quantities $\widetilde{\tau}_k$ have the dimension of time and determine carrier mobility via
\begin{equation}
\label{Eq:mu_dc_bltz_sc_tau_k}
    \mu_\mathrm{dc}^\mathrm{Bltz}=\sum_k v_k^2\widetilde{\tau}_k\frac{e^{-\beta\varepsilon_k}}{Z},
\end{equation}
where $Z=\sum_k e^{-\beta\varepsilon_k}$.
Transforming Eq.~\eqref{Eq:linearized_bltz} as described in Sec.~2.3 of Ref.~\onlinecite{RepProgPhys.83.036501}, we obtain that the quantities $\widetilde{\tau}_k$ have to satisfy the following system of implicit equations:
\begin{equation}
\label{Eq:sc_tau_k_implicit}
    \frac{1}{\widetilde{\tau}_k}=\sum_{q}w_{k+q,k}\left(1-\cos\theta_{k+q,k}\frac{|v_{k+q}|\widetilde{\tau}_{k+q}}{|v_k|\widetilde{\tau}_k}\right),
\end{equation}
where
\begin{equation}
    \cos\theta_{k+q,k}=\frac{v_{k+q}v_k}{|v_{k+q}||v_k|}
\end{equation}
is the cosine of the angle $\theta_{k+q,k}$ between the carrier velocities before and after its scattering on phonons.
In the one-dimensional model we study, $\theta_{k+q,k}$ can take only two values, $0$ and $\pi$.
We emphasize that the ansatz embodied in Eq.~\eqref{Eq:linearized_bltz_ansatz} does not introduce any new approximation to Eq.~\eqref{Eq:linearized_bltz} because it simply restates it as an equation for quantities $\widetilde{\tau}_k$ [Eq.~\eqref{Eq:sc_tau_k_implicit}].
In other words, solving Eq.~\eqref{Eq:sc_tau_k_implicit} for $\widetilde{\tau}_k$ in a self-consistent manner, we obtain the \emph{exact} solution of Eq.~\eqref{Eq:linearized_bltz}, which we compared to the HEOM solution in Fig.~\ref{Fig:heom-vs-boltzmann_071024}.
The data labeled "Boltzmann" in Fig.~\ref{Fig:heom-vs-boltzmann_071024} are obtained by iteratively solving the system in Eq.~\eqref{Eq:sc_tau_k_implicit} starting from
\begin{equation}
\label{Eq:def_serta}
    \frac{1}{\widetilde{\tau}_k^{(0)}}=\frac{1}{\tau_k^\mathrm{SERTA}}=\sum_q w_{k+q,k}.
\end{equation}
This initial guess for $\widetilde{\tau}_k$ is known as the self-energy relaxation-time approximation (SERTA)~\cite{RepProgPhys.83.036501} to the true solution of Eq.~\eqref{Eq:sc_tau_k_implicit}.
The iterative algorithm is stopped once the mobilties computed from Eq.~\eqref{Eq:mu_dc_bltz_sc_tau_k} using the solutions $\widetilde{\tau}_k^{(n-1)}$ and $\widetilde{\tau}_k^{(n)}$ from two consecutive iterations become nearly identical.

\begin{figure}[htbp!]
    \centering
    \includegraphics[width=0.45\columnwidth]{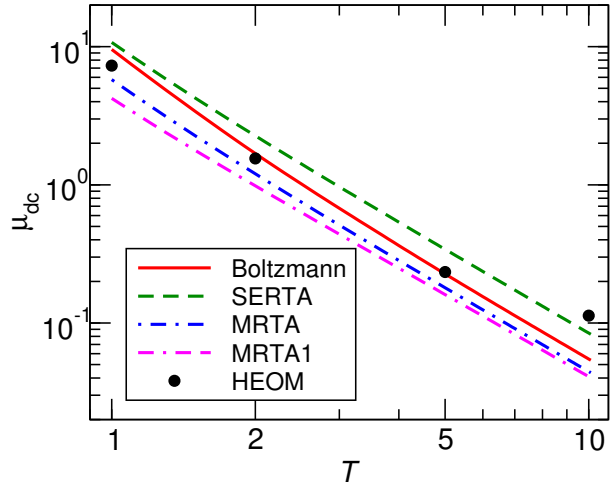}
    \caption{Temperature-dependent mobility computed using the HEOM (symbols) and Eq.~\eqref{Eq:mu_dc_bltz_sc_tau_k} in which $\widetilde{\tau}_k$ is the true (self-consistent) solution to Eq.~\eqref{Eq:sc_tau_k_implicit} (label "Boltzmann") and the approximate solution given in Eq.~\eqref{Eq:def_serta} (label "SERTA"), Eq.~\eqref{Eq:def_mrta} (label "MRTA"), and Eq.~\eqref{Eq:def_mrta1} (label "MRTA1").
    The model parameters are $\omega_0=J=1$ and $\lambda=0.05$.
    The data labeled "Boltzmann", "SERTA", "MRTA", and "MRTA1" are the courtesy of N. Vukmirovi\'c.}
    \label{Fig:boltzmann_appendix}
\end{figure}

Practical computations on first-principles models of real materials~\cite{PhysRevB.97.115203,PhysRevB.104.085203} rely on approximate solutions to Eq.~\eqref{Eq:sc_tau_k_implicit}, such as the SERTA~\cite{PhysRevB.97.115203}.
The momentum relaxation-time approximation (MRTA) assumes that $|v_{k+q}|\widetilde{\tau}_{k+q}\approx|v_k|\widetilde{\tau}_k$ and yields~\cite{PhysRevB.104.085203}
\begin{equation}
\label{Eq:def_mrta}
    \frac{1}{\tau_k^\mathrm{MRTA}}=\sum_q w_{k+q,k}\left(1-\cos\theta_{k+q,k}\right).
\end{equation}
Another widely used version of MRTA (here labeled MRTA1) uses~\cite{RepProgPhys.83.036501,PhysRevResearch.2.013001}
\begin{equation}
\label{Eq:def_mrta1}
    \frac{1}{\tau_k^\mathrm{MRTA1}}=\sum_q w_{k+q,k}\left(1-\frac{v_{k+q}v_k}{|v_k|^2}\right).
\end{equation}
Figure~\ref{Fig:boltzmann_appendix} compares the mobilities in Fig.~\ref{Fig:heom-vs-boltzmann_071024} with the mobilities $\mu_\mathrm{dc}^\mathrm{SERTA}$, $\mu_\mathrm{dc}^\mathrm{MRTA}$, and $\mu_\mathrm{dc}^\mathrm{MRTA1}$ computed by replacing $\widetilde{\tau}_k$ in Eq.~\eqref{Eq:mu_dc_bltz_sc_tau_k} with $\tau_k^\mathrm{SERTA}$, $\tau_k^\mathrm{MRTA}$, and $\tau_k^\mathrm{MRTA1}$, respectively.
We conclude that the widely used approximations to the true solution of Eq.~\eqref{Eq:sc_tau_k_implicit} yield mobilities that either overestimate (SERTA) or underestimate (MRTA and MRTA1) the numerically exact results.
At temperatures $1\leq T/J\leq 5$, when the Boltzmann approach can be justified by the smallness of the phonon-assisted and cross contributions to $\mu_\mathrm{dc}$, we find that the MRTA (MRTA1) underestimates the HEOM results by around 30\% (50\%), while the SERTA overestimates them by around 50\%.
The inaccuracy of the SERTA can be explained by its neglect of the geometric factor that appropriately weighs the contributions from small-angle and large-angle scattering events~\cite{Mahanbook}.
Meanwhile, we ascribe the inaccuracy of MRTA and MRTA1 to the phonon energy being comparable to the carrier energy ($\omega_0/J=1$)~\cite{PhysRevB.104.085203}.
Namely, the approximation $|v_{k+q}|\widetilde{\tau}_{k+q}\approx|v_k|\widetilde{\tau}_k$ underlying MRTA is best satisfied when the change of momentum in a scattering event is small.
This is, however, not the case for $\omega_0=J$, when the relatively large change of carrier's energy $\varepsilon_{k+q}-\varepsilon_k$ suggests that the change $q$ in its momentum is also appreciable.
We expect that the accuracy of the MRTA improves as $\omega_0/J$ is lowered, see also the Supplemental Material of the companion paper~\cite{part2}.
\end{widetext}
\bibliography{aipsamp}
\end{document}


\title{Supplemental Material for\\
Charge transport limited by nonlocal electron--phonon interaction. I. Hierarchical equations of motion approach}

\author{Veljko Jankovi\'c}
 \email{veljko.jankovic@ipb.ac.rs}
 \affiliation{Institute of Physics Belgrade, University of Belgrade, Pregrevica 118, 11080 Belgrade, Serbia}
\maketitle

\section{Inferring the generalized Wick's theorem from the dynamical equations of the HEOM method}
\label{App:GWT_inferred_from_HEOM}
Taking the time derivative of
\begin{equation}
\label{Eq:many-phonon-config}
    \rho_\mathbf{n}^{(n)}(t)=\mathrm{Tr}_\mathrm{ph}\left\{F_\mathbf{n}^{(n)}\rho_\mathrm{tot}(t)\right\},
\end{equation}
and using the Liouville equation $\partial_t\rho_\mathrm{tot}(t)=-i[H_\mathrm{tot},\rho_\mathrm{tot}(t)]$ for the density operator of the interacting carrier--phonon system, one obtains
\begin{equation}
\label{Eq:EOM-from-Liouville}
\begin{split}
    \partial_t\rho_\mathbf{n}^{(n)}(t)
    =&-i[H_\mathrm{e},\rho_\mathbf{n}^{(n)}(t)]\\
    &-i\mathrm{Tr}_\mathrm{ph}\left\{[F_\mathbf{n}^{(n)},H_\mathrm{ph}]\rho_\mathrm{tot}(t)\right\}\\
    &-i\mathrm{Tr}_\mathrm{ph}\left\{F_\mathbf{n}^{(n)}[H_\mathrm{e-ph},\rho_\mathrm{tot}(t)]\right\}.
\end{split}
\end{equation}
In the second term on the RHS of Eq.~\eqref{Eq:EOM-from-Liouville}, we performed a cyclic permutation of phonon operators under the partial trace over phonons.
Inserting $H_\mathrm{e-ph}=\sum_{qm}V_qf_{qm}$ into the third term on the RHS of Eq.~\eqref{Eq:EOM-from-Liouville}, and performing appropriate cyclic permutations of phonon operators, we transform Eq.~\eqref{Eq:EOM-from-Liouville} into
\begin{equation}
\label{Eq:HEOM-from-Liouville}
\begin{split}
    \partial_t\rho_\mathbf{n}^{(n)}(t)
    =&-i[H_\mathrm{e},\rho_\mathbf{n}^{(n)}(t)]\\
    &-i\mathrm{Tr}_\mathrm{ph}\left\{[F_\mathbf{n}^{(n)},H_\mathrm{ph}]\rho_\mathrm{tot}(t)\right\}\\
    &-i\sum_{qm}V_q\mathrm{Tr}_\mathrm{ph}\left\{F_\mathbf{n}^{(n)}f_{qm}\rho_\mathrm{tot}(t)\right\}\\
    &+i\sum_{qm}\mathrm{Tr}_\mathrm{ph}\left\{f_{qm}F_\mathbf{n}^{(n)}\rho_\mathrm{tot}(t)\right\}V_q.
\end{split}
\end{equation}
On the other hand, using $\langle f_{q_2m_2}f_{q_1m_1}\rangle_\mathrm{ph}=\delta_{m_1\overline{m_2}}\eta_{q_2q_1m_2}$ and $\langle f_{q_1m_1}f_{q_2m_2}\rangle_\mathrm{ph}=\delta_{m_1\overline{m_2}}\eta_{\overline{q_2}\:\overline{q_1}\:\overline{m_2}}^*$, we transform the HEOM in Eq.~(5) of the main text into
\begin{equation}
\label{Eq:HEOM-original-1}
\begin{split}
    &\partial_t\rho_\mathbf{n}^{(n)}(t)=-i[H_\mathrm{e},\rho_\mathbf{n}^{(n)}(t)]-\mu_\mathbf{n}\rho_\mathbf{n}^{(n)}(t)\\
    &-i\sum_{qm}V_q\left[\rho_{\mathbf{n}_{qm}^+}^{(n+1)}(t)+\sum_{q'm'}n_{q'm'}\langle f_{q'm'}f_{qm}\rangle_\mathrm{ph}\rho_{\mathbf{n}_{q'm'}^-}^{(n-1)}(t)\right]\\
    &+i\sum_{qm}\left[\rho_{\mathbf{n}_{qm}^+}^{(n+1)}(t)+\sum_{q'm'}n_{q'm'}\langle f_{qm}f_{q'm'}\rangle_\mathrm{ph}\rho_{\mathbf{n}_{q'm'}^-}^{(n-1)}(t)\right]V_q.
\end{split}
\end{equation}
The first terms on the RHSs of Eqs.~\eqref{Eq:HEOM-from-Liouville} and~\eqref{Eq:HEOM-original-1} are identical.
The commutator $[F_\mathbf{n}^{(n)},H_\mathrm{ph}]$ is a purely phononic operator that describes an $n$-phonon-assisted process ($H_\mathrm{ph}$ conserves the number of phonons).
Irrespective of the particular form of $F_\mathbf{n}^{(n)}$, it is clear that the commutator $[F_\mathbf{n}^{(n)},H_\mathrm{ph}]$ is proportional to $F_\mathbf{n}^{(n)}$ itself, thus the second terms of the RHSs of Eqs.~\eqref{Eq:HEOM-from-Liouville} and~\eqref{Eq:HEOM-original-1} have to be identical.
The simplest possibility that the sum of the third and the fourth terms of Eq.~\eqref{Eq:HEOM-from-Liouville} is identical to the sum of the third and the fourth terms of Eq.~\eqref{Eq:HEOM-original-1} is that 
\begin{equation}
\label{Eq:right-1-step-before}
    \mathrm{Tr}_\mathrm{ph}\left\{F_\mathbf{n}^{(n)}f_{qm}\rho_\mathrm{tot}(t)\right\}=\rho_{\mathbf{n}_{qm}^+}^{(n+1)}(t)+\sum_{q'm'}n_{q'm'}\langle f_{q'm'}f_{qm}\rangle_\mathrm{ph}\rho_{\mathbf{n}_{q'm'}^-}^{(n-1)}(t),
\end{equation}
\begin{equation}
\label{Eq:left-1-step-before}
    \mathrm{Tr}_\mathrm{ph}\left\{f_{qm}F_\mathbf{n}^{(n)}\rho_\mathrm{tot}(t)\right\}=\rho_{\mathbf{n}_{qm}^+}^{(n+1)}(t)+\sum_{q'm'}n_{q'm'}\langle f_{qm}f_{q'm'}\rangle_\mathrm{ph}\rho_{\mathbf{n}_{q'm'}^-}^{(n-1)}(t).
\end{equation}
The generalized Wick's theorem embodied in Eqs.~(14) and~(15) of the main text then follows by making use of Eq.~\eqref{Eq:many-phonon-config} on the right-hand sides of Eqs.~\eqref{Eq:right-1-step-before} and~\eqref{Eq:left-1-step-before}, respectively.

\newpage

\section{Implications of the time-reversal symmetry for the cross contribution to the current--current correlation function}
\label{App:time-reversal}
The proof that $\langle j_\mathrm{e}(t)j_\mathrm{e-ph}(0)\rangle=\langle j_\mathrm{e-ph}(t)j_\mathrm{e}(0)\rangle$ relies on general properties of equilibrium correlation functions and the time-reversal operator.

The equilibrium correlation function of hermitean operators $A_2$ and $A_1$ satisfies
\begin{equation}
\label{Eq:general-eq-corr-funct}
    \langle A_2(t)A_1(0)\rangle=\langle A_1(-t)A_2(0)\rangle^*.
\end{equation}
The time-reversal operator $\mathcal{I}_t$ is an antiunitary (antilinear and unitary, $\mathcal{I}_t^{-1}=\mathcal{I}_t^\dagger$), involutive ($\mathcal{I}_t^2={1}$), and thus hermitean ($\mathcal{I}_t^\dagger=\mathcal{I}_t$) operator that acts on the free-electron states $|k\rangle$ as $\mathcal{I}_t|k\rangle=|\overline{k}\rangle$, while its action on phonon creation and annihilation operators is $\mathcal{I}_tb_q^{(\dagger)}\mathcal{I}_t=b_{\overline{q}}^{(\dagger)}$.
The Hamiltonian $H_\mathrm{tot}$ [Eqs.~(1)--(3) of the main text] is invariant under time reversal, i.e., $\mathcal{I}_tH_\mathrm{tot}\mathcal{I}_t=H_\mathrm{tot}$. Using the definition of current operators $j_\mathrm{e}$ [Eq.~(20) of the main text] and $j_\mathrm{e-ph}$ [Eqs.~(21)--(23) of the main text], one obtains that
\begin{equation}
\label{Eq:current-under-I-t}
    \mathcal{I}_t j_\mathrm{e/e-ph}\mathcal{I}_t=-j_\mathrm{e/e-ph}.
\end{equation}
Using the decomposition $\mathcal{I}_t=UK$, where $K$ denotes complex conjugation, while $U$ is a unitary operator, one proves that
\begin{equation}
\label{Eq:similarity-time-reversal}
    \mathrm{Tr}\left\{\mathcal{I}_tA\mathcal{I}_t\right\}=\mathrm{Tr}\left\{A\right\}^*.
\end{equation}

We perform the following transformations
\begin{equation}
\begin{split}
\langle j_\mathrm{e}(t)j_\mathrm{e-ph}(0)\rangle=\langle j_\mathrm{e-ph}(-t)j_\mathrm{e}(0)\rangle^*=\langle\mathcal{I}_tj_\mathrm{e-ph}(-t)\mathcal{I}_t\mathcal{I}_t j_\mathrm{e}\mathcal{I}_t\rangle=(-1)^2\langle j_\mathrm{e-ph}(t)j_\mathrm{e}(0)\rangle.   
\end{split}
\end{equation}
The first equality follows from Eq.~\eqref{Eq:general-eq-corr-funct}.
To establish the second equality, we combine Eq.~\eqref{Eq:similarity-time-reversal} and the invariance of $H_\mathrm{tot}$ under time reversal.
The third equality makes use of the antilinearity of $\mathcal{I}_t$ and Eq.~\eqref{Eq:current-under-I-t}.

\newpage

\section{Details of HEOM computations}
In this section, $N$ denotes the chain length, $D$ is the maximum hierarchy depth, $t_\mathrm{max}$ is the maximum (real) time up to which HEOM are propagated, while $\delta_\mathrm{OSR}$ [Eq.~(51) of the main text] is the relative accuracy with which the optical sum rule is satisfied.
Our HEOM data are openly available in Ref.~\onlinecite{jankovic_2025_14637019}.
\begin{table}[htbp!]
    \centering
    \begin{tabular}{c|c|c|c|c|c|c}
    $\omega_0/J$ & $\lambda$ & $T/J$ & $N$ & $D$ & $t_\mathrm{max}$ & $\delta_\mathrm{OSR}$ \\\hline\hline
    1 & 0.05 & 1 & 160 & 1 & 500 & $9.3\times 10^{-4}$\\\hline
    1 & 0.05 & 2 & 160 & 2 & 300 & $1.15\times 10^{-3}$\\\hline
    1 & 0.05 & 5 & 71 & 3 & 150 & $4\times 10^{-6}$\\\hline
    1 & 0.05 & 10 & 45 & 4 & 100 & $8\times 10^{-5}$\\\hline\hline
    1 & 0.25 & 1 & 45 & 4 & 200 & $10^{-3}$\\\hline
    1 & 0.25 & 2 & 45 & 4 & 70 & $4.5\times 10^{-4}$\\\hline
    1 & 0.25 & 5 & 10 & 7/8 & 100 & $6.6\times 10^{-5}$\\\hline
    1 & 0.25 & $10^{0.8}$ & 10 & 7/8 & 100 & $7.1\times 10^{-5}$\\\hline
    1 & 0.25 & $10^{0.9}$ & 10 & 7/8 & 100 & $7.5\times 10^{-5}$\\\hline
    1 & 0.25 & 10 & 7 & 7/8 & 100 & $8\times 10^{-5}$\\\hline\hline
    1 & 0.5 & 1 & 21 & 6 & 70 & $6\times 10^{-4}$\\\hline
    1 & 0.5 & 2 & 15 & 6 & 100 & $1.9\times 10^{-4}$\\\hline
    1 & 0.5 & 5 & 10 & 7/8 & 50 & $1.2\times 10^{-4}$\\\hline
    1 & 0.5 & $10^{0.8}$ & 10 & 7/8 & 50 & $1.3\times 10^{-4}$\\\hline
    1 & 0.5 & $10^{0.9}$ & 8 & 8/9 & 50 & $1.4\times 10^{-4}$\\\hline
    1 & 0.5 & 10 & 7 & 8/9 & 50 & $1.6\times 10^{-4}$\\\hline\hline
    1 & 1 & 1 & 13 & 8 & 12 & $1.9\times 10^{-3}$\\\hline
    1 & 1 & 2 & 13 & 8 & 15 & $2.1\times 10^{-5}$\\\hline
    1 & 1 & 5 & 9 & 9/10 & 15 & $3.5\times 10^{-4}$\\\hline
    1 & 1 & $10^{0.8}$ & 8 & 10/11 & 15 & $3.9\times 10^{-4}$\\\hline
    1 & 1 & $10^{0.9}$ & 7 & 11/12 & 15 & $4.7\times 10^{-4}$\\\hline
    1 & 1 & 10 & 7 & 11/12 & 10 & $7.0\times 10^{-4}$ 
    \end{tabular}
    \caption{Details of the HEOM computations performed for $\omega_0/J=1$.}
    \label{Tab:details_w0_1}
\end{table}

\begin{table}[htbp!]
    \centering
    \begin{tabular}{c|c|c|c|c|c|c}
    $\omega_0/J$ & $\lambda$ & $T/J$ & $N$ & $D$ & $t_\mathrm{max}$ & $\delta_\mathrm{OSR}$ \\\hline\hline
    3 & 0.05 & 2 & 161 & 2 & 1000 & $2.3\times 10^{-4}$\\\hline
    3 & 0.05 & 5 & 121 & 2 & 400 & $1.0\times 10^{-4}$\\\hline
    3 & 0.05 & 10 & 91 & 2 & 100 & $2.2\times 10^{-4}$\\\hline\hline
    3 & 0.25 & 2 & 31 & 3 & 1000 & $3.6\times 10^{-4}$\\\hline
    3 & 0.25 & 5 & 21 & 5 & 30 & $2.1\times 10^{-4}$\\\hline
    3 & 0.25 & $10^{0.8}$ & 19 & 5 & 30 & $2.2\times 10^{-4}$\\\hline
    3 & 0.25 & $10^{0.9}$ & 17 & 5 & 30 & $2.4\times 10^{-4}$\\\hline
    3 & 0.25 & 10 & 15 & 5 & 30 & $2.5\times 10^{-4}$\\\hline\hline
    3 & 0.5 & 2 & 21 & 5 & 500 & $1.4\times 10^{-4}$\\\hline
    3 & 0.5 & 5 & 15 & 6 & 25 & $2.4\times 10^{-4}$\\\hline
    3 & 0.5 & $10^{0.8}$ & 13 & 6/7 & 25 & $2.6\times 10^{-4}$\\\hline
    3 & 0.5 & $10^{0.9}$ & 10 & 7/8 & 25 & $2.8\times 10^{-4}$\\\hline
    3 & 0.5 & 10 & 10 & 7/8 & 20 & $3.6\times 10^{-4}$\\\hline\hline
    3 & 1 & 2 & 13 & 5 & 500 & $2.9\times 10^{-3}$\\\hline
    3 & 1 & 5 & 13 & 6/7 & 110 & $1.2\times 10^{-4}$\\\hline
    3 & 1 & $10^{0.8}$ & 13 & 6/7 & 110 & $1.2\times 10^{-4}$\\\hline
    3 & 1 & $10^{0.9}$ & 10 & 8/9 & 30 & $2.4\times 10^{-4}$\\\hline
    3 & 1 & 10 & 10 & 8/9 & 20 & $3.7\times 10^{-4}$ 
    \end{tabular}
    \caption{Details of the HEOM computations performed for $\omega_0/J=3$.}
    \label{Tab:details_w0_3}
\end{table}

\newpage

\section{Power-law fits of the temperature-dependent mobility in the regime of phonon-assisted transport}

\begin{figure}[htbp!]
    \centering
    \includegraphics[width=0.5\columnwidth]{fits_w0_1_191224.eps}
    \caption{HEOM results for $\mu_\mathrm{dc}(T)$ (symbols) and their best fits to the power-law function $\mu_\mathrm{dc}(T)=A/T^\alpha$ with two parameters, the amplitude $A$ and the power-law exponent $\alpha$.
    The fits are performed for $\omega_0=J=1$, in parameter regimes in which the phonon-assisted share of the HEOM mobility is $\gtrsim 50\%$ and the magnitude of the cross share is $\lesssim 10\%$, see Figs.~4~(b) and 4~(c) of the main text.
    The values of $\alpha$ are cited next to each dataset.
    Note the logarithmic scale on both axes.
    }
    \label{Fig:fits_w0_1_191224}
\end{figure}

\begin{figure}[htbp!]
    \centering
    \includegraphics[width=0.5\columnwidth]{fits_w0_3_191224.eps}
    \caption{HEOM results for $\mu_\mathrm{dc}(T)$ (symbols) and their best fits to the power-law function $\mu_\mathrm{dc}(T)=A/T^\alpha$ with two parameters, the amplitude $A$ and the power-law exponent $\alpha$.
    The fits are performed for $\omega_0=3$ and $J=1$, in parameter regimes in which the phonon-assisted share of the HEOM mobility is $\gtrsim 50\%$ and the magnitude of the cross share is $\lesssim 10\%$, see Figs.~6~(b) and 6~(c) of the main text.
    The values of $\alpha$ are cited next to each dataset.
    For completeness, we also show HEOM data for $\lambda=0.25$.
    These can be fitted to the power-law function only when the magnitude of the cross contribution falls below $\sim 10\%$, which happens at sufficiently high temperatures, see the red line connecting the last two squares and Fig.~6~(c) of the main text.
    Note the logarithmic scale on both axes.
    }
    \label{Fig:fits_w0_3_191224}
\end{figure}

\newpage

\section{Evaluating the Boltzmann-equation collision integral using the HEOM formalism}
Here, we obtain the collision integral $\left(\frac{\partial p_k}{\partial t}\right)_\mathrm{e-ph}$ for the carrier--phonon scattering in the Boltzmann approach starting from the HEOM.
Taking the matrix element $\langle k|\dots|k\rangle$ of Eq.~(5) of the main text for $\mathbf{n}=0$ and $n=0$, we obtain that the change in the population $p_k(t)=\langle k|\rho(t)|k\rangle$ of the free-carrier state $|k\rangle$ due to the carrier--phonon interaction is
\begin{equation}
\label{Eq:p_k_t_before_elimination}
    \left(\frac{\partial p_k}{\partial t}\right)_\mathrm{e-ph}=-2\sum_{qm}\mathrm{Im}\:\left\{M(k,q)p_{k,qm}^{(1)}(t)\right\},
\end{equation}
where we define
\begin{equation}
    p_{k,qm}^{(1)}(t)=\langle k|\rho_{\mathbf{0}_{qm}^+}^{(1)}(t)|k+q\rangle.
\end{equation}
To arrive at Eq.~\eqref{Eq:p_k_t_before_elimination}, we use $\rho_{\mathbf{0}_{\overline{q}\overline{m}}^+}^{(1)}(t)=\rho_{\mathbf{0}_{qm}^+}^{(1)}(t)^\dagger$.
Taking the matrix element $\langle k|\dots|k+q\rangle$ of Eq.~(5) of the main text for $\mathbf{n}=\mathbf{0}_{qm}^+$ and $n=1$, and neglecting the coupling to HEOM auxiliaries at depth 2, we obtain the following equation for $p_{k,qm}^{(1)}(t)$:
\begin{equation}
\label{Eq:p_k_qm_one_t}
    \partial_t p_{k,qm}^{(1)}(t)=-i\left(\varepsilon_k-\varepsilon_{k+q}-i\mu_m\right)p_{k,qm}^{(1)}(t)-iM(k,q)^*\left[c_m p_{k+q}(t)-c_{\overline{m}}^* p_k(t)\right].
\end{equation}
Integrating Eq.~\eqref{Eq:p_k_qm_one_t} in the Markov approximation $p_k(t-s)\approx p_k(t)$ yields
\begin{equation}
\label{Eq:p_k_qm_one_t_markov}
    p_{k,qm}^{(1)}(t)=-iM(k,q)^*\left[c_m p_{k+q}(t)-c_{\overline{m}}^* p_k(t)\right]\int_0^t ds\:e^{-i(\varepsilon_k-\varepsilon_{k+q}-i\mu_m)s}.
\end{equation}
In the adiabatic approximation, one solves the integral in Eq.~\eqref{Eq:p_k_qm_one_t_markov} by letting $t\to+\infty$ to finally obtain ($\eta\to+0$)
\begin{equation}
\label{Eq:p_k_qm_one_t_markov_adiabatic}
    p_{k,qm}^{(1)}(t)=M(k,q)^*\frac{c_m p_{k+q}(t)-c_{\overline{m}}^* p_k(t)}{\varepsilon_k-\varepsilon_{k+q}-i\mu_m-i\eta}.
\end{equation}
Inserting Eq.~\eqref{Eq:p_k_qm_one_t_markov_adiabatic} into Eq.~\eqref{Eq:p_k_t_before_elimination} and using $c_{\overline{m}}^*=c_{\overline{m}}$ and $\mathrm{Im}\:\frac{1}{\varepsilon_k-\varepsilon_{k+q}-i\mu_m-i\eta}=\pi\delta(\varepsilon_k-\varepsilon_{k+q}-i\mu_m)$ yields the following equation for $p_k(t)$:
\begin{equation}
    \left(\frac{\partial p_k}{\partial t}\right)_\mathrm{e-ph}=-\sum_q w_{k+q,k}p_k(t)+\sum_q w_{k,k+q}p_{k+q}(t),
\end{equation}
where the transition rate from state $|k\rangle$ to state $|k+q\rangle$ is given in Eq.~(E3) of the main text.

\newpage
\bibliography{aipsamp}